\documentclass[conference]{IEEEtran}
\usepackage{cite}
\usepackage{amsmath,amssymb,amsfonts}
\usepackage{algorithmic}
\usepackage{graphicx}
\usepackage{float}
\usepackage{textcomp}
\usepackage{xcolor}
\usepackage{booktabs}
\usepackage{url}
\usepackage{hyperref}
\hypersetup{hidelinks,
  pdftitle={Inference Economics of Enterprise Coding Agents: A Case Study of Cloud vs. On-Premise LLMs},
  pdfauthor={Sheng-Wei Peng, Yi-Hsun Lin, Yi-Pei Lee},
  pdfkeywords={Autonomous Coding Agents, Inference Economics, LLM Quantization (NVFP4), NVIDIA Blackwell, On-Premise LLM Serving, Prompt Caching, Software Repository Mining, Sovereign AI, Total Cost of Ownership}}
\newcommand{\FCR}{\mathrm{FCR}}
\usepackage{verbatim}
\usepackage{tikz}
\usetikzlibrary{shapes.geometric,positioning,patterns}
\usepackage{pgfplots}
\pgfplotsset{compat=1.18}

\begin{document}
\pagestyle{plain}

\title{Inference Economics of Enterprise Coding Agents: A Case Study of Cloud vs.\ On-Premise LLMs}

\author{
\IEEEauthorblockN{Sheng-Wei Peng}
\IEEEauthorblockA{\textit{PEGAVERSE} \\
\textit{Pegatron Corporation}\\
Taipei, Taiwan \\
ken\_peng@pegatroncorp.com}
\and
\IEEEauthorblockN{Yi-Hsun Lin}
\IEEEauthorblockA{\textit{PEGAVERSE} \\
\textit{Pegatron Corporation}\\
Taipei, Taiwan \\
michael\_lin@pegatroncorp.com}
\and
\IEEEauthorblockN{Yi-Pei Lee}
\IEEEauthorblockA{\textit{PEGAVERSE} \\
\textit{Pegatron Corporation}\\
Taipei, Taiwan \\
penny6\_lee@pegatroncorp.com}
}

\maketitle
\thispagestyle{plain}

\begin{abstract}
Autonomous coding agents force engineering organizations to choose between API-based frontier models---strong reasoning at high token cost---and on-premise quantized open-weights models, which promise low-marginal-cost scaling and data sovereignty at some loss of reasoning fidelity. We study this trade-off through a \textit{single-developer, non-randomized} longitudinal case study over two contiguous 28-day periods on a production monorepo: an API-based Claude Opus 4.7/4.8 configuration using \textit{Claude Code} versus an on-premise GLM-5.1/5.2 configuration using \textit{Opencode}, quantized to NVFP4, on NVIDIA Blackwell hardware. Analyzing LLM telemetry and Git history, we find that prompt caching (99.3\% hit rate) cuts realized API cost by 88.6\% to an effective \$0.57 per million tokens---below even the \$2.83 amortized unit cost of the shared on-premise slice (a utilization-dependent inversion; total realized spend and total cost of ownership (TCO) are the robust quantities). At comparable gross code churn, the local configuration was associated with a far higher defect-repair burden: a Fix Commit Ratio ($\FCR$) of 74.9\% versus 45.9\%, with the odds of a commit being a repair 2.6 to 4.9 times higher within every difficulty tier (Mantel--Haenszel $\mathrm{OR} = 3.61$). Under Taiwan-market parameters and a symmetric labor model, on-premise deployment nonetheless saves 40.1\% of true TCO under shared GPU allocation, whereas dedicated reservation costs 43.8\% more than the cached API. Under shared allocation, the genuine penalty is not monetary but a measurable developer-experience burden---timestamp indicators show more work trapped in debugging spirals and a slower commit cadence---and an offline replay shows hybrid routing gateways trade defect rate for infrastructure savings along a cost--quality frontier rather than dominate the pure-API baseline.
\end{abstract}

\begin{IEEEkeywords}
Autonomous Coding Agents, Inference Economics, LLM Quantization (NVFP4), NVIDIA Blackwell, On-Premise LLM Serving, Prompt Caching, Software Repository Mining, Sovereign AI, Total Cost of Ownership.
\end{IEEEkeywords}

\section{Introduction}
Large Language Models (LLMs) have evolved from passive autocomplete utilities (e.g., GitHub Copilot) into autonomous coding agents (e.g., Claude Code, SWE-agent) capable of executing multi-step software engineering workflows. These agents function in closed loops where they inspect file hierarchies, edit source code, invoke compilers or test runners, and dynamically self-correct based on error outputs. The outer loop execution speed and accuracy of such agents are heavily constrained by the reasoning capability, output speed, and token cost structure of their core LLMs.

In enterprise software engineering, two primary archetypes of LLM provision exist:
\begin{enumerate}
    \item \textbf{Commercial Public APIs}: Utilizing closed-source frontier models, such as Claude Opus, hosted by vendors. These models represent the state-of-the-art in reasoning and code generation. However, they require sending code to third-party endpoints and incur token-based pricing.
    \item \textbf{On-Premise Private Clusters}: Deploying open-weights models, such as GLM-5.1/5.2, on internal compute hardware, such as an NVIDIA GB200 NVL72 cluster. These deployments often leverage aggressive quantization techniques (e.g., NVFP4) to achieve high throughput and low latency while keeping code and data within the corporate network~\cite{nvidia2024blackwell}.
\end{enumerate}

To date, few empirical studies have quantified the developer productivity, financial efficiency, and code quality differences between these two paradigms in real-world, large-scale software engineering projects. To bridge this gap, this paper conducts a comparative study of development activities on the production repository of a corporate AI Platform-as-a-Service (PaaS) over two contiguous 28-day periods. During the first period (Period A), the developer utilized Claude Code driven by API-based Claude Opus 4.7/4.8. During the second period (Period B), the developer utilized Opencode driven by GLM-5.1/5.2 NVFP4 running on an internal NVL72 cluster. Methodologically, this is a longitudinal industrial case study: it trades the breadth of a controlled, multi-subject experiment for the depth and ecological validity of continuous production telemetry from a live enterprise codebase; we scope our claims accordingly, deferring generalization to the replication protocol of Section~VII-B.

Through database queries on LLM telemetry and mining of Git commit logs, we provide a detailed analysis of the quantitative token usage, financial expenditure, and software production metrics. Concretely, we investigate four research questions:
\begin{itemize}
    \item \textbf{RQ1 (Cost economics)}: How do the \textit{realized} token economics of a cache-optimized frontier API compare to a self-hosted quantized cluster, both per token and in total? (Sections~V-A and V-D)
    \item \textbf{RQ2 (Code quality)}: How do the two deployments differ in defect-repair burden---Fix Commit Ratio and defect composition---once task difficulty is controlled for? (Sections~V-B and V-C)
    \item \textbf{RQ3 (True TCO)}: Under realistic Taiwan-market operating parameters and a data-grounded labor model, which deployment minimizes the true total cost of ownership (TCO), and how robust is that ordering to parameter choice? (Section~VI-A)
    \item \textbf{RQ4 (Developer experience)}: Beyond direct dollars, what \textit{objective} developer-experience cost does the local configuration impose? (Section~V-G)
\end{itemize}
In answering these, this paper makes the following contributions:
\begin{enumerate}
    \item A ground-truth measurement methodology that triangulates deduplicated LLM telemetry with Git repository mining across two contiguous 28-day production periods, specified in full so it can be reproduced on other repositories and telemetry.
    \item The empirical finding that prompt caching (99.3\% hit rate) inverts the per-token cost comparison between a frontier API and a self-hosted cluster---an effective \$0.57 per million tokens, below even the shared on-premise amortization (\$2.83/M)---a utilization-dependent inversion that reframes the conventional cloud-versus-on-premise cost narrative.
    \item A difficulty-stratified defect analysis ($\mathrm{OR}_{\mathrm{MH}} = 3.61$) demonstrating that the local configuration's defect-repair premium holds within every difficulty tier, independent of a strict output-volume match.
    \item A data-grounded total-cost-of-ownership model under Taiwan-market parameters with a symmetric, timestamp-grounded labor basis and a four-axis sensitivity analysis.
    \item A set of objective, reproducible behavioral workload indicators mined from commit timestamps that quantify the developer-experience cost without subjective self-report.
    \item A cost--quality frontier analysis of hybrid routing gateways via offline counterfactual replay, showing that no routing policy dominates the pure-API baseline; shifting work to the local model trades defect rate for infrastructure savings.
\end{enumerate}

The rest of this paper is organized as follows: Section~II reviews the related work. Section~III contrasts the architectural frameworks of Claude Code and Opencode. Section~IV details the empirical methodology. Section~V presents our quantitative evaluation. Section~VI discusses the strategic business and opportunity costs. Section~VII analyzes threats to validity, and Section~VIII concludes the paper.

\section{Related Work}

\subsection{Autonomous Software Engineering Agents}
Recent work in software engineering has shifted from single-line code prediction toward end-to-end task execution by autonomous agents. Frameworks like SWE-agent~\cite{yang2024sweagent} demonstrate that combining LLMs with a sandboxed shell environment enables agents to locate bugs, modify multiple files, and verify changes via test suites---though simpler non-agentic pipelines remain competitive on the same tasks~\cite{xia2024empirical}. Public benchmarks like SWE-bench~\cite{jimenez2024swebench} evaluate this capacity by testing models on real GitHub issues. However, these agents execute in multi-turn loops, often requiring dozens of sequential model calls to complete a single task. This recursive loop behavior amplifies the impact of underlying model inaccuracies, as minor errors in earlier steps can lead to compounding failures later in the execution trace.

\subsection{Empirical Studies of AI Coding Assistants}
A parallel line of empirical work measures the productivity and quality impact of AI coding assistants on developers. Controlled and field studies of autocomplete-style assistants such as GitHub Copilot report task-completion speedups (e.g., a randomized trial finding a 55.8\% faster completion of a scoped task~\cite{peng2023impact}), alongside more nuanced effects on acceptance behavior and cognitive load. These studies predominantly evaluate \textit{single-suggestion} completion in short lab tasks or self-report surveys; a more recent field randomized controlled trial (RCT) of \textit{agentic} assistants on experienced open-source maintainers found a measured slowdown despite perceived speedup~\cite{metr2025rct}, but reported no cost or deployment dimension. The agentic frameworks above~\cite{yang2024sweagent,jimenez2024swebench} are in turn evaluated mainly on public issue-resolution benchmarks rather than longitudinal production telemetry. Our study complements both lines by mining 56 days of production agent telemetry and Git history, and by contrasting a hosted frontier API against an on-premise quantized deployment---a cost-and-quality dimension absent from prior developer-productivity studies.

\subsection{On-Premise Deployment and Quantization}
For enterprise adoption, on-premise hosting of LLMs has emerged as a viable alternative to commercial APIs. Running models locally requires substantial compute infrastructure. Serving frameworks like vLLM~\cite{kwon2023vllm} and Orca~\cite{yu2022orca} optimize throughput through PagedAttention and iteration-level batching. To optimize hardware efficiency, post-training quantization techniques compress model parameters from 16-bit floating point representations to 8-bit (FP8) or 4-bit (FP4) formats; calibration-based methods such as AWQ~\cite{lin2024awq}, GPTQ~\cite{frantar2023gptq}, and SmoothQuant~\cite{xiao2023smoothquant} reduce the accuracy loss of aggressive quantization. The release of NVIDIA Blackwell GPUs introduced native hardware acceleration for 4-bit formats (NVFP4 and MXFP4)~\cite{nvidia2024blackwell}, and large open-weights Mixture-of-Experts (MoE) models make quantized private serving increasingly practical. While quantization significantly increases token throughput and reduces GPU memory (VRAM) requirements, it can introduce degradation in complex reasoning tasks, such as structural code refactoring and mathematical logic. Vendor-published calibration reports claim near-lossless accuracy for NVFP4 on standard benchmarks~\cite{nvidia2024blackwell,hfglm52}, and benchmark-level studies of quantized code LLMs report mixed degradation~\cite{giagnorio2025quantizing}; the cost on \textit{production agentic} workloads, however, remains uncharacterized, motivating the deployment-level measurement this study provides. The multi-turn loops of Section~II-A then compound even small single-step accuracy losses.

\subsection{Model Selection Matrix and Rationale}
To justify the selection of the local open-weights model and outline the design space, we establish a model selection matrix comparing state-of-the-art open-weights and API-based frontier models available during the study period (leaderboard values accessed 2026-07-07). The evaluation criteria prioritize: (1) context window capacity to prevent retrieval truncation in deep monorepos, (2) on-premise deployability to satisfy enterprise data governance, (3) systems-HPC optimization on Blackwell architectures, and (4) coding intelligence benchmarks.

Table~\ref{tab:model_selection} organizes the design space along the axis that governs on-premise adoption---whether a model can be \textit{self-hosted} at all. \textit{Closed-source frontier} models (Claude Opus, OpenAI's GPT-5.5) are reachable only through a vendor's cloud API and cannot run on internal hardware; \textit{open-weights} models (Zhipu's GLM, DeepSeek, Moonshot AI's Kimi, MiniMax) can. This split makes the endpoints of our study the natural choices on each side. Within the closed-source group, Claude Opus 4.8 leads the compared frontier models on the Artificial Analysis Intelligence Index (56, ahead of GPT-5.5 at its \texttt{xhigh} reasoning setting at 55, and its own predecessor Opus 4.7 at 54)~\cite{artificialanalysis2026}, and it is the model that drives Claude Code. Within the open-weights group, GLM-5.2 is the \textit{highest-ranked open-weights model} on the same index~\cite{artificialanalysis2026} and the top open-weights model on SWE-bench Pro~\cite{morphllm2026}. Selecting these two therefore pits the strongest self-hostable model against the frontier cloud model our agent already used. All Intelligence Index scores are reported at each model's \textit{maximum} reasoning effort, matching our own deployment, which ran both Claude Opus and GLM-5.2 at maximum reasoning effort.

\begin{table*}[htbp]
\caption{Model Selection Matrix for Enterprise Agent Deployment}
\label{tab:model_selection}
\begin{center}
\footnotesize
\setlength{\tabcolsep}{4pt}
\begin{tabular}{llccccc}
\toprule
\textbf{Deployability} & \textbf{Model} & \textbf{Params (Total/Active)} & \textbf{Context} & \textbf{API Price \$/M (in/out)} & \textbf{AA Index$^{\dagger}$} & \textbf{SWE-bench Pro} \\
\midrule
Closed-Source & \textbf{Claude Opus 4.8} & \textbf{Proprietary} & \textbf{1M} & \textbf{\$5.00 / \$25.00} & \textbf{56} & \textbf{69.2\%} \\
(Cloud API only; & GPT-5.5 (xhigh) & Proprietary & 1M & \$5.00 / \$30.00 & 55 & 58.6\% \\
not self-hostable) & Claude Opus 4.7 & Proprietary & 1M & \$5.00 / \$25.00 & 54 & 64.3\% \\
\midrule
Open-Weights & \textbf{GLM-5.2} & \textbf{753B / 40B MoE} & \textbf{1M} & \textbf{\$1.40 / \$4.40} & \textbf{51} & \textbf{62.1\%} \\
(self-hostable & MiniMax M3 & 428B / 23B MoE & 1M & \$0.30 / \$1.20 & 44 & 59.0\% \\
on-premise) & GLM-5.1 & 744B / 40B MoE & 200k & \$1.40 / \$4.40 & 40 & 58.4\% \\
\bottomrule
\end{tabular}
\end{center}
{\footnotesize $^{\dagger}$AA = Artificial Analysis Intelligence Index~\cite{artificialanalysis2026}, reported at each model's \textit{maximum} reasoning effort (e.g., ``max effort''/``xhigh''); SWE-bench Pro scores from the leaderboard aggregation~\cite{morphllm2026}. Each group lists the two model versions used in this study plus the strongest same-class competitor by index; \textbf{bold} marks the primary configuration on each side, both run at maximum reasoning effort. GPT-5.5 price and context are from the vendor's official API documentation~\cite{openaipricing2026}.}
\end{table*}

The model selection matrix reveals key architectural trade-offs:
\begin{itemize}
    \item \textbf{Coding Capability and Openness}: At the time of deployment, GLM-5.2 held the highest SWE-bench Pro score among open-weights models (62.1\%, versus 59.0\% for MiniMax M3)~\cite{morphllm2026}, and ranked as the top open-weights model on the Artificial Analysis Intelligence Index (51, versus 56 for Claude Opus 4.8)~\cite{artificialanalysis2026}. Combined with its MIT license and a native 1M-token context window, this made it the strongest candidate for private deployment.
    \item \textbf{Serving and Memory Efficiency}: GLM-5.2 pairs a 753B total-parameter MoE structure with only 40B active parameters per token, and NVIDIA publishes an official NVFP4 quantization checkpoint with an FP8 Key-Value (KV) cache~\cite{hfglm52}. This permits serving on a four-GPU slice---one GB200 NVL72 compute tray---rather than the $\ge$8-GPU allocation required for unquantized checkpoints, yielding a cost-efficient private deployment. Larger open-weights alternatives such as DeepSeek-V4 Pro impose materially higher VRAM floors.
    \item \textbf{API Cost Optimization and Prompt Caching}: Claude Opus 4.8 carries the highest nominal pricing among the models deployed in this study (\$5.00 input, \$25.00 output per million tokens). However, Anthropic's prompt caching discounts cached prefix reads by 90\% (to \$0.50 per million tokens), and cache-heavy agentic workloads realize most input tokens as cache reads (99.3\% in our telemetry, Section~V). This caching mechanism renders Claude far more price-competitive in repetitive multi-turn development workflows than its nominal rates suggest, while the on-premise cost basis remains hardware-hours rather than tokens.
\end{itemize}
This architectural decision matrix justifies the selection of GLM-5.2 (and its predecessor GLM-5.1 prior to the 5.2 release on 2026-06-13) as the local benchmark model, balancing context capacity, hardware overhead, and coding intelligence.

\subsection{Prompt Caching in Agentic Workflows}
Autonomous coding agents are highly context-intensive, as they frequently send substantial parts of the codebase, JSDocs, database schemas, and tool definitions with every request. Prompt caching (e.g., Anthropic's Prompt Caching) allows the inference server to reuse the KV cache of static prompt prefixes. This skips the redundant prefill computation, yielding significant latency reductions (up to 85\%) and input cost discounts (up to 90\%)~\cite{anthropic2024promptcache}. In agentic frameworks, the prefix remains highly static (containing system instructions and repository index), making such workloads ideal candidates for prompt caching.

\section{System Architectures: Claude Code vs.\ Opencode}
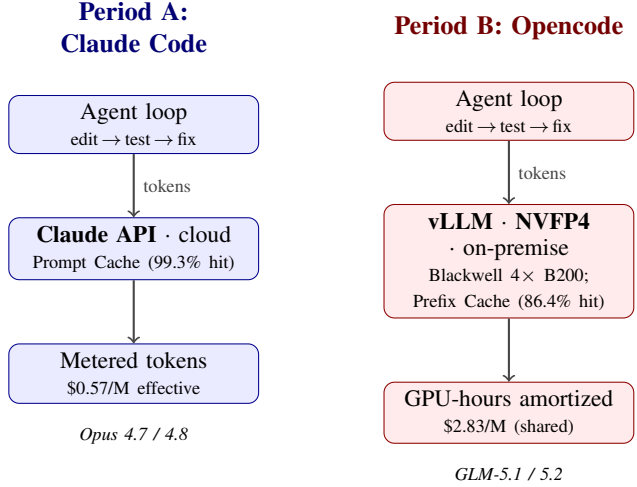
\begin{figure}[htbp]
\centering
\resizebox{0.98\columnwidth}{!}{
\begin{tikzpicture}[node distance=0.45cm, font=\small,
  cloud/.style={rectangle, draw, rounded corners, align=center, text width=3.15cm, minimum height=0.7cm, inner sep=3pt, fill=blue!8, draw=blue!55!black},
  onprem/.style={rectangle, draw, rounded corners, align=center, text width=3.15cm, minimum height=0.7cm, inner sep=3pt, fill=red!8, draw=red!55!black},
  ar/.style={->, thick, gray!55!black},
]
\node[font=\bfseries, text=blue!45!black, text width=3.3cm, align=center] (hA) {Period A: Claude Code};
\node[font=\bfseries, text=red!45!black, text width=3.3cm, align=center, right=1.5cm of hA] (hB) {Period B: Opencode};
\node[cloud, below=of hA] (loopA) {Agent loop\\ \scriptsize edit\,$\rightarrow$\,test\,$\rightarrow$\,fix};
\node[onprem, below=of hB] (loopB) {Agent loop\\ \scriptsize edit\,$\rightarrow$\,test\,$\rightarrow$\,fix};
\node[cloud, below=0.85cm of loopA] (beA) {\textbf{Claude API} $\cdot$ cloud\\ \scriptsize Prompt Cache (99.3\% hit)};
\node[onprem, below=0.85cm of loopB] (beB) {\textbf{vLLM $\cdot$ NVFP4} $\cdot$ on-premise\\ \scriptsize Blackwell 4$\times$ B200; Prefix Cache (86.4\% hit)};
\node[cloud, below=0.85cm of beA] (coA) {Metered tokens\\ \scriptsize \$0.57/M effective};
\node[onprem, below=0.85cm of beB] (coB) {GPU-hours amortized\\ \scriptsize \$2.83/M (shared)};
\node[font=\scriptsize\itshape, below=0.2cm of coA] {Opus 4.7 / 4.8};
\node[font=\scriptsize\itshape, below=0.2cm of coB] {GLM-5.1 / 5.2};
\draw[ar] (loopA)-- node[right,font=\scriptsize]{tokens} (beA);
\draw[ar] (loopB)-- node[right,font=\scriptsize]{tokens} (beB);
\draw[ar] (beA)--(coA);
\draw[ar] (beB)--(coB);
\end{tikzpicture}
}
\caption{Deployable-configuration architectures compared in this study. \textbf{Period A} pairs Claude Code with the hosted Claude API; \textbf{Period B} pairs Opencode with a self-hosted vLLM engine serving NVFP4-quantized GLM on a Blackwell 4$\times$ B200 tray, priced by amortized GPU-hours. The model axis (Opus vs.\ GLM) and deployment axis (cloud vs.\ on-premise) are the two contrasted variables; cache-hit rates and unit prices preview results from Section~V.}
\label{fig:architecture}
\end{figure}

A critical aspect of this study is comparing the software architectures of the two agent frameworks: \textit{Claude Code} and \textit{Opencode}. Although both serve as terminal-based autonomous agents, they are built on fundamentally different engineering paradigms, contrasted at a glance in Fig.~\ref{fig:architecture}.

\subsection{Claude Code Execution Model}
Claude Code is a commercial, Node.js-based terminal agent developed by Anthropic~\cite{claudecode2026}. It operates locally, interacting directly with the filesystem and local shell commands, and handles task execution by issuing API calls to the hosted Anthropic service. The local client parses the repository directory and dynamically updates the context window with file contents, terminal execution output, and history of past agent steps, which is then submitted to the remote inference models.

\subsection{Opencode Framework Architecture}
Opencode is an open-source (MIT-licensed) terminal coding agent~\cite{opencode2026}. It follows a client--server design: a background server process manages session state, conversation history, and filesystem operations, while a terminal-UI (TUI) client connects to it, so a session persists across terminal restarts. The server core is written in TypeScript and executed on the Bun runtime, whereas the TUI client is implemented in Go; model access is mediated through a provider-abstraction layer (the Vercel AI SDK), making the agent provider-agnostic. In our deployment, Opencode drove the on-premise Blackwell cluster serving the quantized GLM models via an OpenAI-compatible endpoint, with every session in its full-tool-access \textit{Build Mode}.

Unlike the single-vendor architecture of Claude Code, Opencode's design centers on provider neutrality:
\begin{itemize}
    \item \textbf{Bun Runtime Core}: The server core runs on Bun rather than Node.js, providing rapid cold-start times and native TypeScript execution; the interactive TUI client is a separate Go process.
    \item \textbf{Local Session Persistence}: Sessions, message history, and token-usage metrics are persisted in a local SQLite database managed through the Drizzle ORM, independent of any vendor service.
    \item \textbf{Provider-Agnostic Model Access}: Through the Vercel AI SDK provider abstraction, the framework binds provider-specific system-prompt profiles and tool-schema encodings at session start (e.g., an Anthropic-tuned profile versus generic profiles for open-weights backends), rather than assuming a single vendor's prompt conventions.
\end{itemize}

\subsection{Architectural Trade-offs}
Table~\ref{tab:arch_comparison} highlights the design contrasts between the two agent frameworks.

\begin{table}[htbp]
\caption{Architectural Comparison of Agent Frameworks}
\label{tab:arch_comparison}
\begin{center}
\footnotesize
\setlength{\tabcolsep}{3pt}
\begin{tabular}{lll}
\toprule
\textbf{Dimension} & \textbf{Claude Code} & \textbf{Opencode} \\
\midrule
Runtime Engine & Node.js & Bun \\
Source Availability & Closed-source & Open-source (MIT) \\
Session State & JSON transcripts & SQLite (Drizzle ORM) \\
Agent Interface & CLI / IDE extensions & CLI $+$ terminal UI \\
Prompt Profiles & Vendor-tuned & Per-provider profiles \\
Model Coupling & Anthropic models only & Provider-agnostic \\
Prompt Caching & Server-side KV reuse & vLLM prefix caching \\
\bottomrule
\end{tabular}
\end{center}
\end{table}

Crucially, these engineering dimensions act as capability multipliers for the underlying LLMs, explaining a portion of the performance differences observed in practice:
\begin{itemize}
    \item \textbf{Vendor Co-design}: Claude Code and the Claude Opus models are developed by the same vendor, and the harness (tool schemas, system prompts, edit formats) is co-tuned with the model's training. Opencode, as a provider-agnostic framework, must adapt heterogeneous models through generic prompt profiles, which cannot exploit model-specific alignment to the same degree.
    \item \textbf{Prompt Caching Integration}: Claude Code automatically inserts cache breakpoints so that static prefixes (system rules, tool schemas, conversation history) are reused server-side. Our local vLLM deployment similarly benefits from automatic prefix caching (86.4\% of Period B prompt tokens served as cache hits, Section~V-A).
\end{itemize}
These engineering differences represent a construct confound, discussed further in Section~VII-C.

\subsection{Base Model Performance Benchmarks}
To understand the underlying capabilities of the models driving the autonomous coding agents in our evaluation, we compile independently aggregated benchmark and market data. Table~\ref{tab:model_benchmarks} compares the commercial cloud models (Claude Opus 4.8 and 4.7) against the local open-weights Mixture-of-Experts models (GLM-5.2 and 5.1) using the Artificial Analysis Intelligence Index~\cite{artificialanalysis2026}, the SWE-bench Pro leaderboard~\cite{morphllm2026}, and measured serving throughput of the vendors' hosted endpoints.

\begin{table}[htbp]
\caption{Base Model Comparison (Independently Aggregated)}
\label{tab:model_benchmarks}
\begin{center}
\footnotesize
\setlength{\tabcolsep}{2.5pt}
\begin{tabular}{lrrrr}
\toprule
\textbf{Metric} & \textbf{Opus 4.8} & \textbf{Opus 4.7} & \textbf{GLM-5.2} & \textbf{GLM-5.1} \\
\midrule
AA Intelligence Index & 56 & 54 & 51 & 40 \\
SWE-bench Pro & 69.2\% & 64.3\% & 62.1\% & 58.4\% \\
Hosted output (tok/s) & 66.0 & 56.6 & 214.9 & 76.4 \\
Price \$/M (in/out) & \$5/\$25 & \$5/\$25 & \$1.40/\$4.40 & \$1.40/\$4.40 \\
Context window & 1M & 1M & 1M & 200k \\
Release date & 2026-05-28 & 2026-04-16 & 2026-06-13 & 2026-04-07 \\
\bottomrule
\end{tabular}
\end{center}
\end{table}

The aggregated scores show that GLM-5.2 was the strongest open-weights coding model at the time of the study---ranked first among open-weights models on both the Intelligence Index and SWE-bench Pro---yet a persistent gap remains against the frontier API models (Intelligence Index 51 vs.\ 56; SWE-bench Pro 62.1\% vs.\ 69.2\%). Our empirical evaluation examines how this single-digit benchmark gap, compounded by NVFP4 quantization and provider-agnostic harness adaptation, is associated with a much larger gap in end-to-end defect rates ($\FCR$ 74.9\% vs.\ 45.9\%, Section~V).

\subsection{Unified Multi-Dimensional Ecosystem Comparison}
To provide a complete comparative view of the two paradigms, we analyze the multi-dimensional differences between the hosted API model (Opus) and the on-premise open-weights cluster (GLM), alongside their corresponding CLI frameworks (Claude Code and Opencode).

\subsubsection{Infrastructure Footprint and Scaling Limits}
Claude Code acts as a thin client requiring negligible local CPU and memory. However, it is bound by vendor-side, per-organization rate limits (request- and token-per-minute quotas). In contrast, Opencode requires an on-premise private GPU cluster (in our deployment, four NVIDIA Blackwell B200-class GPUs---one GB200 NVL72 compute tray) to serve the quantized GLM model. While this demands substantial capital expenditure and VRAM allocation (approximately 465~GB of NVFP4 weights plus an FP8 KV cache), it removes external rate limits, allowing parallel requests across the corporate intranet up to the cluster's serving capacity.

\subsubsection{Context Window Dynamics and Latency Profiles}
The core models exhibit distinct context capabilities. Claude Opus 4.7 and 4.8 feature a 1,000,000-token context window, allowing the Claude Code agent to ingest large directory structures and multi-turn conversation history in a single prompt, with prefill latency amortized via server-side prompt caching. On the local side, the deployed checkpoints differ by generation: the GLM-5.1 NVFP4 checkpoint is configured for a 202,752-token context, whereas the GLM-5.2 NVFP4 checkpoint supports 1,048,576 tokens (verified from the deployed Hugging Face model configurations~\cite{hfglm51,hfglm52}). Local serving via vLLM likewise enabled automatic prefix caching (86.4\% hit, Section~V-A); because the local cost basis is hardware-hours (Section~V-D), the TCO and defect-rate results are unaffected by the caching configuration.

\subsubsection{Data Governance and Network Compliance}
Data residency is a critical distinction. Operating Claude Code involves transmitting corporate intellectual property (source code, database schema, logs) to Anthropic's cloud endpoints. This requires compliance audits under data protection regulations (such as GDPR). Opencode, operating within a private Kubernetes namespace on an internal network, keeps inference traffic inside the corporate intranet, which substantially simplifies compliance for proprietary technology assets relative to third-party API transmission.

\section{Methodology}

\subsection{Development Environment and Tools}
This study was conducted on a production codebase of a corporate AI PaaS monorepo consisting of 12 distinct packages (including frontends in TypeScript/React and backends in Python/FastAPI). 
\begin{itemize}
    \item \textbf{Period A (Claude Code + Claude Opus)}: From 2026-05-11 to 2026-06-07. The developer utilized the \textit{Claude Code} terminal agent, interacting with commercial API endpoints of Claude Opus 4.7 (released 2026-04-16) and, after its release on 2026-05-28, Claude Opus 4.8. Telemetry attributes 28,615 requests to Opus 4.7 and 3,286 to Opus 4.8.
    \item \textbf{Period B (Opencode + GLM)}: From 2026-06-08 to 2026-07-05. The developer transitioned to \textit{Opencode}, backed by private NVFP4-quantized deployments on a shared NVIDIA Blackwell cluster: GLM-5.1 for the initial days (GLM-5.2 was released on 2026-06-13), followed by GLM-5.2 served first from a community NVFP4 checkpoint and later from NVIDIA's official NVFP4 checkpoint. Both periods therefore span a mid-period upgrade to the vendor's latest generation, keeping the ``rolling frontier'' condition symmetric. Although GLM-5.1 and GLM-5.2 differ more in capability (Intelligence Index 40 vs.\ 51) than Opus 4.7 and 4.8 (54 vs.\ 56), the weekly defect series (Fig.~\ref{fig:weekly_fcr}) shows no structural break at these checkpoint transitions, so Period B is analyzed as a single configuration, symmetric with Period A's pooling of Opus 4.7/4.8.
\end{itemize}

\subsection{Data Collection}
\begin{figure}[htbp]
\centering
\resizebox{0.92\columnwidth}{!}{
\begin{tikzpicture}[node distance=0.5cm, font=\small,
  b/.style={rectangle, draw, rounded corners, align=center, text width=4.3cm, minimum height=0.6cm, inner sep=3pt, fill=gray!7},
  side/.style={rectangle, draw, dashed, rounded corners, align=center, text width=2.5cm, minimum height=0.55cm, inner sep=2pt, font=\footnotesize},
  tel/.style={rectangle, draw, rounded corners, align=center, text width=3.0cm, minimum height=0.6cm, inner sep=2pt, font=\footnotesize, fill=gray!20, draw=gray!55!black},
  fin/.style={rectangle, draw, rounded corners, align=center, text width=4.3cm, minimum height=0.6cm, inner sep=3pt, fill=green!9, draw=green!55!black},
  ar/.style={->, thick, gray!55!black},
]
\node[b] (g1) {Git: all branches $+$ PR refs (\texttt{--no-merges})};
\node[b, below=of g1] (g2) {\texttt{patch-id} dedup (earliest author timestamp)};
\node[b, below=of g2] (g3) {Keyword classify: $\FCR$ $+$ defect taxonomy};
\node[b, below=of g3] (g4) {Difficulty tiering: High / Med / Low};
\node[fin, below=of g4] (m) {Metrics: statistics / cost / behavior};
\node[side, left=of g2] (rm) {Redmine cross-validation};
\node[tel, right=of m] (tel) {Telemetry: Langfuse (Period A) $+$ vLLM engine (Period B)};
\draw[ar] (g1)--(g2); \draw[ar] (g2)--(g3); \draw[ar] (g3)--(g4); \draw[ar] (g4)--(m);
\draw[ar, dashed] (rm)--(g2);
\draw[ar] (tel)--(m);
\end{tikzpicture}
}
\caption{End-to-end measurement pipeline: Git history (all branches and PR refs) is \texttt{patch-id}-deduplicated, Redmine-cross-validated, then keyword-classified and difficulty-stratified; token telemetry (Langfuse for Period A, vLLM engine counters for Period B) feeds the cost metrics. All rules are deterministic (Sections~IV-B--IV-E).}
\label{fig:methodology}
\end{figure}
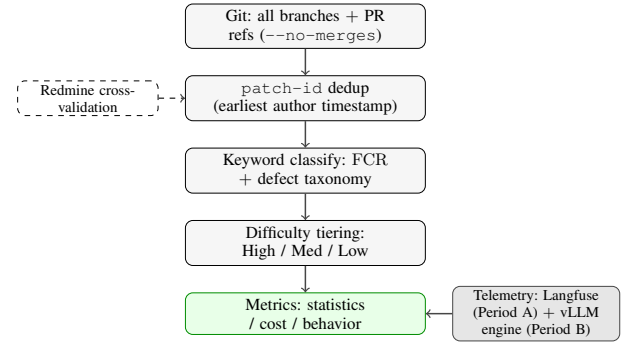

We gathered data from two primary channels:
\begin{enumerate}
    \item \textbf{Telemetry Logs}: Period A's hosted Claude API telemetry was recorded in a Langfuse server deployed in a private namespace on Kubernetes, and extracted directly from its ClickHouse database backend~\cite{langfuse2025telemetry} on 2026-07-07 by aggregating observation-level usage records (\texttt{input}, \texttt{output}, \texttt{cache\_creation\_input\_tokens}, \texttt{cache\_read\_input\_tokens}) grouped by model and filtered to the developer's user identifier and to the coding-agent models (\texttt{claude-opus-4-7} and \texttt{claude-opus-4-8}); auxiliary model traffic from unrelated pipelines is excluded. Two extraction details are essential for correctness: (i) because Langfuse's ClickHouse tables use a \texttt{ReplacingMergeTree} engine, queries deduplicate multi-version rows via \texttt{FINAL} with \texttt{is\_deleted = 0} on both the observations and traces tables (omitting this inflates request counts by roughly 3\% in our data); and (ii) period boundaries are evaluated in the developer's local time zone (\texttt{Asia/Taipei}, UTC+8), matching the time zone semantics of the Git mining, rather than the server's native UTC. These two conditions apply to the Langfuse-recorded Period A telemetry. Period B's local GLM traffic, by contrast, was served by an on-premise vLLM stack, whose engine-level Prometheus counters (prompt, cached, and recomputed tokens) provide the Period B token and cache accounting, with prefix-cache hits reported directly (86.4\%); these source the Period B column of Table~\ref{tab:tokens}. The two periods thus use the backend-appropriate telemetry surface---Langfuse for the hosted Claude API, vLLM engine metrics for the local GLM cluster---not a single shared pipeline.
    \item \textbf{Git Repository Mining}: Git logs from the PaaS repository were parsed to isolate the development commits under study via their Git author identity, from which we extracted line additions, deletions, files changed, and keyword-classified bug-fix commits~\cite{mockus2000identifying}; all commit-level metrics exclude merge commits (\texttt{--no-merges}). To capture the developer's authored work completely---including commits developed on feature branches or rebased prior to merging---we enumerate every non-merge commit authored by the developer over each period across all branches and pull-request refs. Two safeguards keep this from over-counting: \texttt{git patch-id} collapses rebased or cherry-picked copies of the same change to a single instance (retaining the earliest authoring timestamp), and each commit is assigned to a period solely by its author timestamp, so only in-window work under the developer's identity is counted. As an additional attribution check, the mined commits were cross-validated against the team's Redmine issue tracker (Table~\ref{tab:sdlc}), corroborating---without adjusting---the author-and-timestamp filter; every metric is computed directly from Git. We further verified that every mined work stream was ultimately merged into the mainline through the reviewed pull-request flow (Section~IV-G): the commit series measures accepted, review-approved output---including the repair iterations that precede acceptance---rather than abandoned drafts. Figure~\ref{fig:methodology} summarizes the end-to-end extraction and reconstruction pipeline across both channels.
\end{enumerate}

\subsection{Defect Metric Formulation}
To evaluate code quality, we define the \textit{Fix Commit Ratio} ($\FCR$) as:
\begin{equation}
\FCR = \frac{C_{\text{fix}}}{C_{\text{total}}} \times 100\%,
\end{equation}
where $C_{\text{fix}}$ represents the count of commits whose message matches the regular expression pattern \texttt{/(fix|bug|issue|patch|correct)/i}, indicating that the commit was dedicated to repairing defects, resolving compile/lint errors, or fixing broken tests~\cite{mockus2000identifying}. $C_{\text{total}}$ is the total number of commits by the developer during the period.

\subsubsection{Classification Reproducibility}
Both the fix/non-fix decision and the defect taxonomy assignment (Section~V-B) are produced by deterministic keyword rules applied to commit subjects, rather than ad-hoc human judgment: the fix decision uses the regular expression above, and taxonomy assignment applies a fixed priority order of category-specific keyword sets (test-related, then syntax/type-related, then API/dependency-related, with the remainder classified as logical defects). The complete rule set is specified in this paper, making every classification decision reproducible from a repository's own commit history. The first author manually reviewed \textit{every} classified commit and confirmed its rule assignment; because this rater is not blind to the hypothesis, we recommend a blind second-rater protocol reporting inter-rater agreement (Cohen's $\kappa$) as a replication check. The residual risk is construct-level---which changes should count as defect repairs at all---and is discussed under construct validity (Section~VII-C).

\subsection{Economic and Feedback Loop Model}
To evaluate the true cost of using autonomous coding agents, we construct an economic model adapted from classical software engineering economics~\cite{boehm1981software}, modeling both compute expenses and developer hourly costs. Let the total corporate cost of development ($TC$) for $M$ tasks be represented as:
\begin{equation}
TC = M \times \left( C_{\text{compute}} + C_{\text{ops}} + T_{\text{iter}} \times R_{\text{dev}} \right),
\end{equation}
where $C_{\text{compute}}$ is the compute/inference cost per task, $C_{\text{ops}}$ is the operational infrastructure overhead per task (inclusive of electricity, cooling, facilities, and personnel support), $R_{\text{dev}}$ is the developer's hourly wage rate, and $T_{\text{iter}}$ is the feedback loop iteration time required to successfully complete a single coding task. 

We formalize $T_{\text{iter}}$ as a function of developer time and model performance:
\begin{equation}
T_{\text{iter}} = T_{\text{coding}} + N_{\text{round}} \times \left( T_{\text{wait}} + T_{\text{review}} \right),
\end{equation}
where $T_{\text{coding}}$ is the developer's initial scoping and coding time, $T_{\text{wait}}$ is the average inference wait time per request (determined by model throughput and network latency), $T_{\text{review}}$ is the developer's cognitive review and debugging time per iteration, and $N_{\text{round}}$ is the average number of interaction rounds. A local on-premise model achieves a net financial advantage over a commercial API model if and only if:
\begin{equation}
\Delta C_{\text{compute}} > R_{\text{dev}} \times \Delta T_{\text{iter}},
\end{equation}
where $\Delta C_{\text{compute}} = C_{\text{compute, api}} - C_{\text{compute, local}}$ and $\Delta T_{\text{iter}} = T_{\text{iter, local}} - T_{\text{iter, api}}$.

\subsection{Task Difficulty and Category Mapping}
To assess task selection effects, we classify every non-merge commit in both periods by category and difficulty using deterministic rules computed from the Git metadata itself. Category follows the conventional-commit prefix (\texttt{feat}, \texttt{fix}, \texttt{refactor}/\texttt{perf}/\texttt{style}, \texttt{test}/\texttt{ci}, \texttt{docs}/\texttt{chore}). Difficulty is derived from the change footprint: High (touching $\ge 3$ packages, modifying $\ge 5$ files, or $>200$ net lines), Medium (touching $2$ packages or modifying $2$--$4$ files), and Low (single-file modifications). Both rules are fully reproducible from the repository history.

Table~\ref{tab:difficulty} displays the resulting distribution. Unlike an idealized controlled experiment, the two periods are \textit{not} homogeneous: Period A contains a larger share of High-difficulty commits (35.0\% vs.\ 23.4\%; $\chi^2(2) = 13.14$, $p = 0.0014$, Cram\'{e}r's $V = 0.15$), reflecting that Period A included heavier feature and infrastructure work while Period B skewed toward smaller changes. We address this confound in two ways. First, all defect comparisons in Section~V are additionally reported \textit{stratified by difficulty tier}, and the local configuration exhibits a higher fix-commit share within every tier. Second, the direction of the imbalance is conservative with respect to our conclusion: the period that handled \textit{harder} tasks (Period A, Claude) produced the \textit{lower} defect-repair share, so equalizing difficulty would widen, not narrow, the observed gap.

\begin{table}[htbp]
\caption{Commit Category and Difficulty Distribution (non-merge commits)}
\label{tab:difficulty}
\begin{center}
\footnotesize
\setlength{\tabcolsep}{3pt}
\begin{tabular}{llrrrr}
\toprule
\textbf{Period} & \textbf{Category} & \textbf{High} & \textbf{Med} & \textbf{Low} & \textbf{Total} \\
\midrule
Period A & Feature & 28 & 6 & 3 & 37 \\
(Opus) & Refactor/Perf/Style & 7 & 5 & 6 & 18 \\
& Fix (\texttt{fix:} prefix)$^{\ddagger}$ & 8 & 39 & 54 & 101 \\
& Test/CI & 12 & 4 & 11 & 27 \\
& Docs/Chore & 31 & 13 & 19 & 63 \\
& \textbf{Total} & \textbf{86} & \textbf{67} & \textbf{93} & \textbf{246} \\
& \textit{Tier share} & \textit{35.0\%} & \textit{27.2\%} & \textit{37.8\%} & \\
\midrule
Period B & Feature & 31 & 14 & 4 & 49 \\
(GLM) & Refactor/Perf/Style & 8 & 7 & 3 & 18 \\
& Fix (\texttt{fix:} prefix)$^{\ddagger}$ & 34 & 110 & 111 & 255 \\
& Test/CI & 3 & 4 & 6 & 13 \\
& Docs/Chore & 10 & 9 & 13 & 32 \\
& \textbf{Total} & \textbf{86} & \textbf{144} & \textbf{137} & \textbf{367} \\
& \textit{Tier share} & \textit{23.4\%} & \textit{39.2\%} & \textit{37.3\%} & \\
\bottomrule
\end{tabular}
\end{center}
{\footnotesize $^{\ddagger}$The \textit{Fix} \textit{category} here partitions commits by their conventional-commit prefix (\texttt{fix:}) and is disjoint from the other categories; it is distinct from the \textit{Fix Commit Ratio} count of Section~IV-C (113 for Period A, 275 for Period B), which applies the regex \texttt{/(fix|bug|issue|patch|correct)/i} to the full commit message and therefore also captures repair commits filed under other prefixes. The per-tier $\FCR$ analysis (Section~V-C) uses the latter (regex) count. Percentages may not sum to 100 due to rounding.}
\end{table}

\subsection{Specification-Driven Development (SDD) Protocol}
Both evaluation periods utilized a Specification-Driven Development (SDD) methodology to guide the autonomous agents. In the SDD workflow, the developer first creates a detailed markdown specification file (e.g., \texttt{feature\_spec.md} or OpenAPI schemas) outlining the required functions, database schemas, and type definitions. The agent is then executed with the specification file injected directly into its context.

The agent's task execution loop follows a strict specification-compliance protocol: (1) parsing the specifications, (2) generating or updating corresponding TypeScript interfaces and backend schemas, (3) creating unit tests, and (4) implementing the core logic. While the Opus configuration generally produced type-safe implementations matching the specification in a few iterations, the GLM configuration more frequently violated the specified type constraints and API boundaries, yielding type-check and runtime integration failures that required developer intervention---a difference we quantify in Section~V-B.

\subsection{Software Quality-Assurance and Delivery Pipeline}
Both evaluation periods were conducted not on a throwaway prototype but inside the enterprise software development lifecycle (SDLC) of the production PaaS monorepo, under a single, uniformly enforced quality-assurance and delivery pipeline. This property is central to the study's internal and construct validity: every commit we mine---in either period---is a \textit{gated} artifact admitted through the \textit{same} multi-stage quality gate and, once merged, promoted along the \textit{same} multi-environment path. Because the pipeline is held fixed while the coding agent, model, quantization, and serving stack vary together, engineering-process maturity is controlled by design and cannot manufacture a between-period difference; the $\FCR$ and defect deltas of Section~V therefore isolate the deployable configuration rather than the surrounding engineering discipline. Table~\ref{tab:sdlc} summarizes the controls.

The quality gate is a single source of truth---one command, invoked identically by every runner (workstation, mandatory pre-push hook, nightly full run)---spanning linting, strict type checking, layered tests behind a per-file coverage gate, migration-chain checks, a full software-supply-chain security suite, and load/benchmark/chaos testing (detailed in Table~\ref{tab:sdlc}). Merges require human review under CODEOWNERS and Conventional-Commit enforcement plus---notably for this study---two AI review agents served by an internally hosted model, so the team dogfoods on-premise agentic inference within its own review loop. Releases follow near-daily semantic-versioned delivery, promoting one Helm chart across six environment overlays with rolling updates and automatic rollback. We report this pipeline not as a contribution in itself but as evidence that the mined commits reflect production-grade, quality-gated engineering under both configurations.

\begin{table*}[t]
\caption{Software Development Lifecycle Controls Applied Uniformly Across Both Evaluation Periods (Section~IV-G)}
\label{tab:sdlc}
\centering
\footnotesize
\begin{tabular}{@{}l p{9.4cm} p{4.6cm}@{}}
\toprule
\textbf{SDLC stage} & \textbf{Controls and tooling} & \textbf{Enforcement point} \\
\midrule
Requirements \& design & Specification-driven development (versioned \texttt{specs/}), architecture decision records, and design documents; Redmine issue\,$\leftrightarrow$\,PR\,$\leftrightarrow$\,commit traceability & Specification precedes implementation \\
\addlinespace[2pt]
Coding standards & Ruff, \texttt{mypy} ($\times5$ projects), ESLint/\texttt{tsc}; a written team style guide and an agent-instruction file; Conventional Commits & Pre-push and commit-msg Git hooks \\
\addlinespace[2pt]
Code review & $\ge1$ human approval under CODEOWNERS with a PR / Definition-of-Done template; two AI review agents served by an internally hosted model & Branch protection; required status checks \\
\addlinespace[2pt]
Test & \texttt{pytest}/\texttt{vitest} behind a per-file coverage gate, E2E, OpenAPI contract snapshots, and DB-migration chain checks; load, benchmark, and chaos suites & Pre-push (fast subset) $+$ nightly (full) \\
\addlinespace[2pt]
Security \& supply chain & SAST (Bandit, Semgrep, CodeQL), secret scanning (gitleaks), dependency audit (pip/npm), image and filesystem CVE scans (Trivy, grype), SBOM (syft), and an SPDX license allowlist; threat modeling, penetration testing, and a secret manager with rotation & Nightly gate; periodic security audit \\
\addlinespace[2pt]
Build \& release & Single-source-of-truth CI (one command, every runner); semantic versioning with automated changelog; continuous, near-daily delivery & Git-hook enforced \\
\addlinespace[2pt]
Deployment & One Helm chart across six environment overlays (\texttt{local}, \texttt{standalone}, \texttt{test}, \texttt{staging}, \texttt{production}, plus a shared-PVC variant); Helm lint, \texttt{kubeconform}, and dry-run validation; rolling update with automatic rollback & Strict promotion; manual production approval \\
\addlinespace[2pt]
Runtime \& ops & Metrics, centralized logging, distributed tracing, and alerting; on-call rotation, post-incident reviews, and defined SLOs & Continuous monitoring \\
\bottomrule
\end{tabular}
\end{table*}

\section{Empirical Evaluation}

\subsection{Token Usage and Caching Characteristics}
Table~\ref{tab:tokens} presents the token telemetry for both periods, restricted to the coding-agent models (Period A from the Langfuse ClickHouse backend per Section~IV-B; Period B from the on-premise vLLM engine's Prometheus counters, whose request counts are not directly comparable to Langfuse's).

\begin{table}[htbp]
\caption{Token Telemetry Comparison. Period B (vLLM) meters only cache-hit vs.\ compute; all computed prefill is cached, hence reported as Cache Creation, with no separate incremental input (---).}
\label{tab:tokens}
\begin{center}
\footnotesize
\begin{tabular}{lrr}
\toprule
\textbf{Metric} & \textbf{Period A (Opus)} & \textbf{Period B (GLM)} \\
\midrule
Total Requests & 31,901 & 12,281 \\
Incremental Input Tokens & 980,326 & --- \\
Cache Creation Tokens & 108,060,378 & 122,472,214 \\
Cache Read Tokens & 15,202,019,495 & 778,058,773 \\
Total Output Tokens & 20,156,852 & 5,059,772 \\
\textbf{Total Tokens} & \textbf{15,331,217,051} & \textbf{905,590,759} \\
\midrule
Cache Hit Rate (\%) & \textbf{99.3\%} & \textbf{86.4\%} \\
\bottomrule
\end{tabular}
\end{center}
\end{table}

Under the agentic loop of Claude Code, every request re-transmits the accumulated session context. This results in an enormous raw prompt volume of \textbf{15.31 billion} input-side tokens (approximately 480k prompt tokens per request) for Opus. However, due to Anthropic's prompt caching mechanism, \textbf{99.3\%} of these prompt tokens were served as cache reads; only \textbf{0.7\%} were billed at uncached rates. The two output volumes also differ sharply: the Opus configuration generated 20.2M output tokens versus 5.06M for GLM---a 4.0$\times$ difference. Since the two periods delivered comparable code volume (Section~V-B), this reflects Claude Code's more verbose agentic loop (extensive reasoning traces and repeated full-file rewrites) rather than a code-output gap.

The GLM deployment exhibits the same reliance on prefix caching: with vLLM's automatic prefix caching enabled, Period B served 86.4\% of its prompt tokens (778M of 900M) as KV-cache hits, with the recomputed-prefill counter effectively zero; the agentic loop's prompt prefixes are equally repetitive. In our cost model (Section~V-D), caching reduces Claude's realized API cost by 88.6\% relative to the same token volume at nominal input rates, consistent with the up-to-90\% discount bound documented for prefix caching in agentic workflows~\cite{anthropic2024promptcache}.

\subsection{Software Output and Code Quality Metrics}
Table~\ref{tab:git} summarizes the Git mining results from the PaaS repository.

\begin{table}[htbp]
\caption{Git Productivity and Quality Metrics (non-merge commits)}
\label{tab:git}
\begin{center}
\footnotesize
\begin{tabular}{lrr}
\toprule
\textbf{Metric} & \textbf{Period A (Opus)} & \textbf{Period B (GLM)} \\
\midrule
Total Commits ($C_{\text{total}}$) & 246 & 367 \\
Files Changed & 2,469 & 2,159 \\
Insertions ($L_{\text{ins}}$) & 129,556 & 93,281 \\
\quad of which code-only$^{\dagger}$ & 75,738 & 80,258 \\
Deletions ($L_{\text{del}}$) & 38,838 & 34,608 \\
Net Line Growth ($L_{\text{net}}$) & \textbf{90,718} & \textbf{58,673} \\
\midrule
Fix Commits ($C_{\text{fix}}$) & 113 & 275 \\
Fix Commit Ratio ($\FCR$) & \textbf{45.93\%} & \textbf{74.93\%} \\
\bottomrule
\end{tabular}
\end{center}
{\footnotesize $^{\dagger}$Excluding Markdown documentation, dependency lockfiles, and extension-less files.}
\end{table}

\begin{figure}[htbp]
\centering
\begin{tikzpicture}
\begin{axis}[
    width=\columnwidth, height=5.95cm,
    tick label style={font=\footnotesize},
    label style={font=\footnotesize},
    xlabel={Day of Period},
    ylabel={Cumulative Net LOC},
    xmin=1, xmax=28,
    ymin=0, ymax=100000,
    xtick={1, 4, 8, 12, 16, 20, 24, 28},
    scaled y ticks=false,
    ytick={0, 20000, 40000, 60000, 80000, 100000},
    yticklabels={0,20k,40k,60k,80k,100k},
    legend pos=south east,
    grid=both,
    grid style={line width=.1pt, draw=gray!10},
    major grid style={line width=.2pt, draw=gray!20},
    legend style={font=\scriptsize},
]
\addplot[thick, color=blue!70!black, mark=square*, mark options={fill=blue!40}] coordinates {
    (1, 167) (4, 853) (8, 77558) (12, 80628) (16, 86697) (20, 86423) (24, 89912) (28, 90718)
};
\addplot[thick, color=red!70!black, mark=*, mark options={fill=red!40}] coordinates {
    (1, 0) (4, 10292) (8, 15172) (12, 19346) (16, 35861) (20, 36772) (24, 39794) (28, 58673)
};
\legend{Period A (Opus), Period B (GLM)}
\end{axis}
\end{tikzpicture}
\caption{Cumulative net line growth over the 28-day periods (daily Git data). Period A concentrates most growth in a large feature burst during week 2; Period B grows more steadily, including a mid-period feature push around days 15--21.}
\label{fig:loc_growth}
\end{figure}
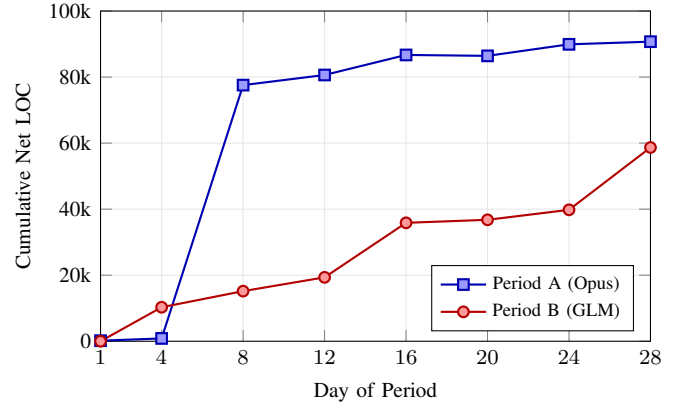

\begin{table}[htbp]
\caption{Inserted Lines by Language / File Type}
\label{tab:languages}
\begin{center}
\footnotesize
\setlength{\tabcolsep}{3.5pt}
\begin{tabular}{lrr}
\toprule
\textbf{Language / File Type} & \textbf{Period A (Opus)} & \textbf{Period B (GLM)} \\
\midrule
Python (\texttt{.py}) & 48,666 (37.6\%) & 38,813 (41.6\%) \\
Markdown docs (\texttt{.md}) & 47,405 (36.6\%) & 9,381 (10.1\%) \\
TypeScript (\texttt{.ts}, \texttt{.tsx}) & 22,151 (17.1\%) & 35,593 (38.2\%) \\
Config \& Shell$^{\S}$ & 4,921 (3.8\%) & 5,804 (6.2\%) \\
Other & 6,413 (4.9\%) & 3,690 (4.0\%) \\
\midrule
\textbf{Total Insertions ($L_{\text{ins}}$)} & \textbf{129,556 (100\%)} & \textbf{93,281 (100\%)} \\
\bottomrule
\end{tabular}
\end{center}
{\footnotesize $^{\S}$\texttt{.json}, \texttt{.sh}, \texttt{.yaml}, \texttt{.toml}. Percentages may not sum to 100 due to rounding.}
\end{table}

Commit counts were higher in Period B (367 vs.\ 246 non-merge commits). Total inserted volume was higher in Period A (129,556 vs.\ 93,281 lines), but this gap is largely an artifact of documentation: 36.6\% of Period A's insertions were Markdown specification files produced under the SDD protocol (Section~IV-F), versus only 10.1\% in Period B (Table~\ref{tab:languages}). \textit{Restricted to code-only insertions, the two periods are comparable---indeed Period B is marginally higher (80,258 vs.\ 75,738 lines).} We take code-only gross insertions as the volume-invariance basis. The all-file \textit{net} growth of Table~\ref{tab:git} runs the other way (90,718 vs.\ 58,673, favoring Period A), again a documentation artifact compounded by Period A's larger deletion churn---which only strengthens the quality reading: Period A produced more retained net lines \textit{and} carried the lower defect-repair share. The two configurations therefore delivered a \textit{similar gross code churn} (Fig.~\ref{fig:loc_growth} traces the cumulative net-line trajectory of each period); because the all-file net growth diverges, we do not claim strict output-volume invariance and instead rest the quality comparison on the difficulty-stratified within-tier analysis ($\mathrm{OR}_{\mathrm{MH}} = 3.61$, Section~V-C), which does not depend on equal volume. However, raw LOC is a flawed proxy for developer productivity when accompanied by high defect rates. To measure the actual utility of the generated code, we define the \textit{Effective Productivity Ratio} as $E_{\text{net}} = 1 - \FCR$, the proportion of commits that deliver new logic rather than defect corrections. With an $\FCR$ of $45.93\%$, Period A retains $E_{\text{net}}^A = 54.07\%$ of its commits as net-new logic; with an $\FCR$ of $74.93\%$ (roughly three out of four commits were defect corrections), Period B retains only $E_{\text{net}}^B = 25.07\%$.

This effective-yield gap restates the measured $\FCR$ difference: under the local GLM setup, the developer's activity was dominated by repairing agent-generated defects rather than writing new logic. $E_{\text{net}}$ is a coarse discount---treating every fix commit as zero-value---but it bounds the direction and rough magnitude of the quality adjustment (Section~V-G).

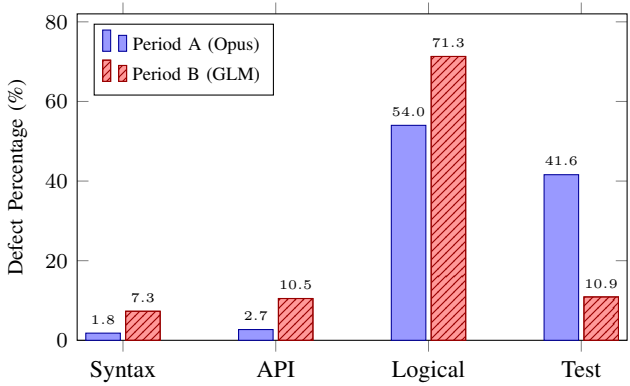
\begin{figure}[htbp]
\centering
\begin{tikzpicture}
\begin{axis}[
    ybar,
    bar width=13pt,
    width=\columnwidth,
    height=5.9cm,
    ylabel={Defect Percentage (\%)},
    label style={font=\footnotesize},
    y tick label style={font=\footnotesize},
    symbolic x coords={Syntax, API, Logical, Test},
    xtick=data,
    ymin=0, ymax=82,
    legend pos=north west,
    legend style={font=\scriptsize},
    x tick label style={font=\small},
    nodes near coords,
    nodes near coords style={font=\tiny, /pgf/number format/precision=1, /pgf/number format/fixed, /pgf/number format/fixed zerofill},
]
\addplot[fill=blue!40, draw=blue!60!black] coordinates {
    (Syntax, 1.8)
    (API, 2.7)
    (Logical, 54.0)
    (Test, 41.6)
};
\addplot[fill=red!40, draw=red!60!black, postaction={pattern=north east lines, pattern color=red!60!black}] coordinates {
    (Syntax, 7.3)
    (API, 10.5)
    (Logical, 71.3)
    (Test, 10.9)
};
\legend{Period A (Opus), Period B (GLM)}
\end{axis}
\end{tikzpicture}
\caption{Percentage distribution of defect categories classified from fix commits. Period A's repair load is dominated by logical and test/CI-infrastructure work, whereas Period B shifts markedly toward syntax/type and API/dependency defects.}
\label{fig:defect_distribution}
\end{figure}

To further analyze the quality gap, Table~\ref{tab:defects} provides a defect taxonomy classifying all non-merge fix commits during both periods, using the deterministic keyword rules of Section~IV-C. The distributions differ, but the difference is concentrated in the Test/CI category---a Period-A infrastructure confound (Fig.~\ref{fig:defect_distribution}); we test its statistical significance in Section~V-C.

\subsubsection{Syntax and Type Discipline}
The GLM configuration produced 20 syntax/type-repair commits versus 2 under Opus (7.3\% vs.\ 1.8\% of fix commits---a $10\times$ absolute inflation). A representative Period B example is a static-typing violation in which the agent assigned an endpoint response to a variable without the required \texttt{dict | list} union annotation, failing \texttt{mypy} on the exception branch (Listing~1). We note that message-level mining underestimates this category for both configurations: type errors that the agent repairs within the same working session never surface as separate fix commits, so these counts reflect only the residual type errors that escaped the agent's own iteration loop and required dedicated repair commits.

\subsubsection{Dependency Mapping and API Boundary Compliance}
API, import, and dependency misuse accounted for 10.5\% of GLM's fix commits (29 commits) versus 2.7\% for Opus (3 commits)---a nearly $10\times$ absolute inflation. Notably, this gap cannot be attributed to context capacity: the deployed GLM-5.2 checkpoint supports the same 1M-token context class as Claude Opus (Section~III-E). The failure modes instead concern the correct use of repository-internal and library APIs: hallucinated import paths, stale method signatures, migration-revision collisions, and environment-behavior misuse such as an HTTP client inheriting proxy environment variables inside the cluster network (Listing~2). This suggests that long-context capacity alone does not confer accurate API grounding in a large monorepo.

\subsubsection{Logical Defects Dominate the GLM Repair Load}
Logical defects are the largest category in both periods but dominate under GLM: 196 commits (71.3\% of fix commits) versus 61 (54.0\%) under Opus---a $3.2\times$ absolute inflation. Period B logical repairs include broken control flow around failure paths (e.g., aborting swap operations on deleted resources, preserving recoverable state on post-commit failures) that required developer-directed correction after the agent's initial implementation shipped subtly incorrect behavior.

\subsubsection{Test and CI Repair: An Infrastructure Confound in Period A}
The test/CI category inverts the pattern: 47 commits (41.6\%) under Opus versus 30 (10.9\%) under GLM. Inspection shows that much of Period A's repair volume was test-infrastructure maintenance (raising CI memory limits, scoping coverage guards, deflaking suites) coinciding with a phase of active CI build-out, rather than repairs of model-generated logic. This has an important interpretive consequence: Period A's $\FCR$ of 45.9\% partially reflects infrastructure churn unrelated to model output quality, meaning the raw $\FCR$ comparison likely \textit{understates} the model-attributable quality gap between the two configurations. We revisit this as a construct-validity consideration in Section~VII-C.

\begin{table}[htbp]
\caption{Defect Taxonomy Comparison (keyword-rule classification)}
\label{tab:defects}
\begin{center}
\footnotesize
\begin{tabular}{lrr}
\toprule
\textbf{Defect Category} & \textbf{Period A (Opus)} & \textbf{Period B (GLM)} \\
\midrule
Syntax \& Type Errors & 2 (1.8\%) & 20 (7.3\%) \\
API \& Package Misuse & 3 (2.7\%) & 29 (10.5\%) \\
Logical Defects & 61 (54.0\%) & 196 (71.3\%) \\
Test \& CI Repairs & 47 (41.6\%) & 30 (10.9\%) \\
\midrule
\textbf{Total Fix Commits} & \textbf{113 (100\%)} & \textbf{275 (100\%)} \\
\bottomrule
\end{tabular}
\end{center}
{\footnotesize Percentages may not sum to 100 due to rounding.}
\end{table}

\subsection{Statistical Hypothesis Testing}
To validate the productivity and quality differences observed between the two periods, we formulate three statistical hypotheses and perform non-parametric Mann--Whitney U tests (Wilcoxon rank-sum) on the Git-derived metrics, since daily software metrics do not conform to a normal distribution. We measure effect sizes using Cliff's Delta ($\Delta$)~\cite{kitchenham2024recommendations}, a non-parametric measure recommended for small-sample software-engineering experiments, classified as negligible ($|\Delta| < 0.147$), small ($|\Delta| < 0.33$), medium ($|\Delta| < 0.474$), or large ($|\Delta| \geq 0.474$).

The three hypotheses are defined as follows:
\begin{itemize}
    \item $H_0^{(1)}$: There is no difference in daily commit frequency between Period A and Period B.
    \item $H_0^{(2)}$: There is no difference in code volume per unit of work between Period A and Period B.
    \item $H_0^{(3)}$: There is no difference in daily $\FCR$ between Period A and Period B.
\end{itemize}

Table~\ref{tab:statistical_testing} summarizes the results.

\begin{table}[htbp]
\caption{Statistical Hypothesis Testing Results (non-merge commits)}
\label{tab:statistical_testing}
\begin{center}
\footnotesize
\setlength{\tabcolsep}{2pt}
\begin{tabular}{lrrrl}
\toprule
\textbf{Metric} & \textbf{A (Opus)} & \textbf{B (GLM)} & \textbf{$p$} & \textbf{Cliff's $\Delta$} \\
\midrule
Commits / active day (mean) & 11.18 & 18.35 & 0.066 & $-0.332$ (M) \\
Insertions / active day (mean) & 5,889 & 4,664 & 0.801 & $-0.046$ (N) \\
Insertions / commit (median) & 36 & 25 & 0.068 & $+0.087$ (N) \\
Daily FCR, active days (mean) & 38.1\% & 72.8\% & $0.0002$ & $-0.661$ (L) \\
\bottomrule
\end{tabular}
\end{center}
{\footnotesize Daily metrics are means over active days (22 in Period A, 20 in Period B). Cliff's $\Delta > 0$ indicates Period A stochastically larger; effect sizes: N = negligible, M = medium, L = large (thresholds of Section~V-C).}
\end{table}

Three findings emerge. First, we fail to reject $H_0^{(1)}$ at $\alpha = 0.05$ ($p = 0.066$): daily commit count does not differ significantly across periods, though Period B trends toward a higher count (mean 18.4 vs.\ 11.2, $\Delta = -0.332$). Second, for $H_0^{(2)}$ we report a transparent null result: neither the day-level insertion distribution ($p = 0.801$) nor the per-commit insertion size (median 36 vs.\ 25 lines; $p = 0.068$) differs significantly, with negligible effect sizes---consistent with the comparable code volume (75.7k vs.\ 80.3k code-only lines) that holds output roughly constant. Third, $H_0^{(3)}$ is rejected decisively ($p = 0.0002$) with a large effect size ($\Delta = -0.66$): on comparable working days, the share of repair work under the local GLM configuration is dramatically higher.

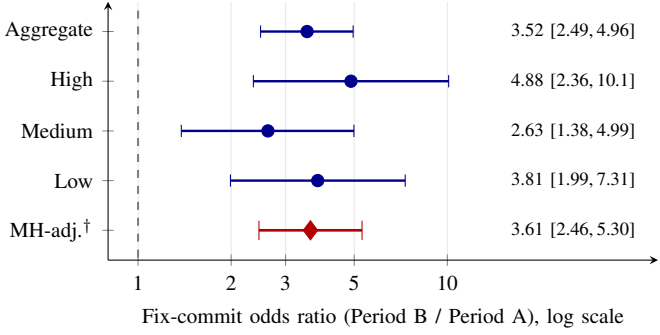
\begin{figure}[htbp]
\centering
\begin{tikzpicture}
\begin{axis}[
    width=\columnwidth,
    height=5.0cm,
    xmode=log,
    log basis x=10,
    xmin=0.8, xmax=48,
    ymin=0.4, ymax=5.6,
    xtick={1,2,3,5,10},
    xticklabels={1,2,3,5,10},
    ytick={1,2,3,4,5},
    yticklabels={MH-adj.$^{\dagger}$, Low, Medium, High, Aggregate},
    xlabel={Fix-commit odds ratio (Period B / Period A), log scale},
    tick label style={font=\footnotesize},
    xlabel style={font=\footnotesize},
    axis lines=left,
    xmajorgrids, grid style={gray!18},
    clip=false,
]
\addplot[dashed, gray, thick, forget plot] coordinates {(1,0.45) (1,5.55)};
\addplot[only marks, mark=*, mark size=2.3pt, color=blue!55!black,
    error bars/.cd, x dir=both, x explicit, error bar style={line width=0.9pt, color=blue!55!black}]
    coordinates {
    (3.52,5) += (1.44,0) -= (1.03,0)
    (4.88,4) += (5.22,0) -= (2.52,0)
    (2.63,3) += (2.36,0) -= (1.25,0)
    (3.81,2) += (3.50,0) -= (1.82,0)
    };
\addplot[only marks, mark=diamond*, mark size=3.6pt, color=red!70!black,
    error bars/.cd, x dir=both, x explicit, error bar style={line width=1pt, color=red!70!black}]
    coordinates {
    (3.61,1) += (1.69,0) -= (1.15,0)
    };
\node[anchor=west, font=\scriptsize] at (axis cs:15,5) {3.52 [2.49,\,4.96]};
\node[anchor=west, font=\scriptsize] at (axis cs:15,4) {4.88 [2.36,\,10.1]};
\node[anchor=west, font=\scriptsize] at (axis cs:15,3) {2.63 [1.38,\,4.99]};
\node[anchor=west, font=\scriptsize] at (axis cs:15,2) {3.81 [1.99,\,7.31]};
\node[anchor=west, font=\scriptsize] at (axis cs:15,1) {3.61 [2.46,\,5.30]};
\end{axis}
\end{tikzpicture}
\caption{Forest plot of the fix-commit odds ratio (Period B $/$ Period A), overall and stratified by task-difficulty tier. Markers are point estimates; whiskers are 95\% CIs (Wald on the log-odds; Robins--Breslow--Greenland for the Mantel--Haenszel summary, $^{\dagger}$). Every interval lies entirely right of the null (OR $=1$, dashed): the local GLM configuration carries a significantly higher defect-repair share \textit{within every difficulty tier}, not merely in aggregate (Table~\ref{tab:statistical_testing}, Section~V-C).}
\label{fig:odds_forest}
\end{figure}

To quantify the association on the aggregated commit contingency table (Period A: 113 fix vs.\ 133 non-fix; Period B: 275 fix vs.\ 92 non-fix), a Pearson $\chi^2$ test of independence rejects the null hypothesis with high significance: $\chi^2(1) = 53.30$, $p < 0.0001$, $\phi = 0.295$. The unadjusted Odds Ratio is $3.52$ (95\% CI $[2.49, 4.96]$). Because the two periods differ in difficulty composition (Section~IV-E), we additionally compute the Mantel--Haenszel odds ratio stratified by difficulty tier, obtaining $\mathrm{OR}_{\mathrm{MH}} = 3.61$, with per-tier odds ratios of $4.88$ (High: 46.5\% vs.\ 15.1\%), $2.63$ (Medium: 80.6\% vs.\ 61.2\%), and $3.81$ (Low: 86.9\% vs.\ 63.4\%). The stratified and unadjusted estimates are close, demonstrating that the defect-repair premium of the local configuration is not an artifact of task mix: within every difficulty tier, the odds of a commit being a defect repair are $2.6$--$4.9\times$ higher under GLM (Fig.~\ref{fig:odds_forest}). We flag one construct nuance: because difficulty is derived from change footprint and fix commits are systematically smaller (median 25 vs.\ 36 lines, fix vs.\ non-fix), the Low tier is mechanically enriched with fixes (hence Claude's inverted ordering: Low 63.4\% $>$ Medium 61.2\% $>$ High 15.1\%). This couples the stratifier to the outcome \textit{base rate}, not to the \textit{between-period} contrast; applied identically to both periods, it cannot manufacture the odds-ratio gap---so the tiers read as change-size strata rather than intrinsic-difficulty measures.

Finally, the defect-taxonomy distributions of Table~\ref{tab:defects} differ significantly between deployments ($\chi^2(3) = 51.93$, $p < 0.0001$, Cram\'{e}r's $V = 0.37$); however, this difference is driven by the Test/CI category, whose Period-A excess is infrastructure churn unrelated to model output (Section~V-B, ``Test and CI Repair''). Excluding Test/CI, the remaining model-attributable categories do not differ significantly ($\chi^2(2) = 5.59$, $p = 0.06$, $V = 0.13$), so the two deployments differ primarily in the \textit{volume} of repair work, not its structural composition.

We treat the daily-FCR difference ($H_0^{(3)}$) as primary and the rest as descriptive; the principal results survive multiplicity. Under a conservative Bonferroni correction across the seven tests here---the four contrasts of Table~\ref{tab:statistical_testing}, the aggregate and taxonomy $\chi^2$ tests, and the Mantel--Haenszel stratified contrast ($\alpha = 0.05/7 = 0.0071$)---the daily-FCR ($p = 0.0002$), aggregate-fix ($p < 0.0001$), and defect-taxonomy ($p < 0.0001$) differences all remain significant, while the non-significant tests are unaffected. Other tests reported elsewhere (per-language FCR, weekly trends) are exploratory and excluded from this confirmatory family. No conclusion depends on a $p$-value near threshold.

\textit{Scope of inference.} These tests take the day or the individual commit as the unit of analysis. Because every observation derives from one developer on one repository, these units are \textit{not} independent replicates: consecutive days and commits are autocorrelated, and the design is single-subject and two-period rather than randomized between-subjects. The reported $p$-values and CIs therefore quantify how strongly and consistently the difference manifests \textit{within this developer and project}, and do not by themselves license inference to a broader population; we defer that to the multi-developer cross-over protocol of Section~VII-B.

\subsection{Financial Cost Evaluation}
We calculate the financial cost of both configurations. For the Claude API, we apply Anthropic's published Opus 4.7/4.8 rates with prompt caching (\$5.00/M base input, \$6.25/M cache write, \$0.50/M cache read, \$25.00/M output)~\cite{anthropicpricing2026}. The on-premise workload ran on a shared NVIDIA GB200 NVL72 cluster; for cost accounting we attribute it to the \textit{minimum viable serving footprint} of GLM-5.2 NVFP4---a 4$\times$ B200 slice, one GB200 NVL72 compute tray (Section~V-H), i.e., a lower bound on the hardware a team must provision. We evaluate two rental-equivalent scenarios for this footprint at \$16.00/hour (\$4.00 per GPU-hour, consistent with June--July 2026 market on-demand rates of \$3.70--\$5.90 per B200-hour across major GPU clouds~\cite{b200pricing2026}):
\begin{itemize}
    \item \textbf{Scenario I (Dedicated Reservation)}: Dedicated around-the-clock allocation for the full period (24~h $\times$ 28~d = 672 hours) = \$10,752.00 USD.
    \item \textbf{Scenario II (Shared Allocation)}: Allocation charged as a standard 160-hour monthly developer seat under the shared cluster's chargeback model = \$2,560.00 USD.
\end{itemize}
Because both agent configurations ran around the clock---including unattended sessions---per-developer attribution is a billing convention rather than a measurement; Scenarios I and II bracket that attribution spectrum, from full wall-clock reservation to a standard seat.

\begin{table}[htbp]
\caption{Financial Cost Comparison (USD; Scenarios I/II = 4$\times$ B200 Dedicated/Shared)}
\label{tab:cost}
\begin{center}
\footnotesize
\setlength{\tabcolsep}{3pt}
\begin{tabular}{lrrr}
\toprule
\textbf{Item} & \textbf{Claude API} & \textbf{Scenario I} & \textbf{Scenario II} \\
\midrule
Input Cost & \$4.90 & --- & --- \\
Cache Write & \$675.38 & --- & --- \\
Cache Read & \$7,601.01 & --- & --- \\
Output Cost & \$503.92 & --- & --- \\
\midrule
\textbf{Total Cost} & \textbf{\$8,785.21} & \textbf{\$10,752.00} & \textbf{\$2,560.00} \\
\midrule
Compute Savings vs.\ API$^{\dagger}$ & --- & \textbf{$-22.4\%$} & \textbf{70.9\%} \\
Eq. Cost/Million Tokens & \$0.573 & \$11.87 & \$2.83 \\
\bottomrule
\end{tabular}
\end{center}
{\footnotesize $^{\dagger}$Total-bill comparison; the \textit{per-token} rates below invert it---the cached API is cheaper per processed token despite the larger bill---as analyzed in Section~V-D.}
\end{table}

Two structural observations follow from Table~\ref{tab:cost}. First, absent caching, the same Claude token volume would have cost \$77,059 at nominal input rates; prompt caching therefore reduced realized spend by 88.6\%. Second, and counterintuitively, caching drives Claude's \textit{effective unit price} (\$0.57 per million processed tokens) \textit{below} both the shared on-premise amortization (\$2.83/M) and the dedicated-reservation unit cost (\$11.87/M). The API's total bill remains higher only because the Claude configuration processed $16.9\times$ more tokens: per processed token, the cached frontier API is the \textit{cheaper} option---an inversion of what nominal price sheets suggest. The on-premise per-token figure divides a time-billed resource by one developer's token volume, so it is inflated by low single-tenant utilization; we therefore treat total realized spend and the TCO of Section~VI-A as the robust quantities, with the per-token rate as corroboration.

Extrapolating to a 100-developer organization, a pure API approach would cost approximately \$879,000 USD monthly versus \$256,000 USD for shared on-premise allocation. We caution that this assumes the per-developer amortization holds at fleet scale: serving 100 concurrent workloads needs proportionally more GPU capacity (one 4$\times$ B200 tray cannot serve 100 developers), so the extrapolation compares amortization \textit{rates}, not a fixed footprint.

\subsection{Model Performance Benchmarking}
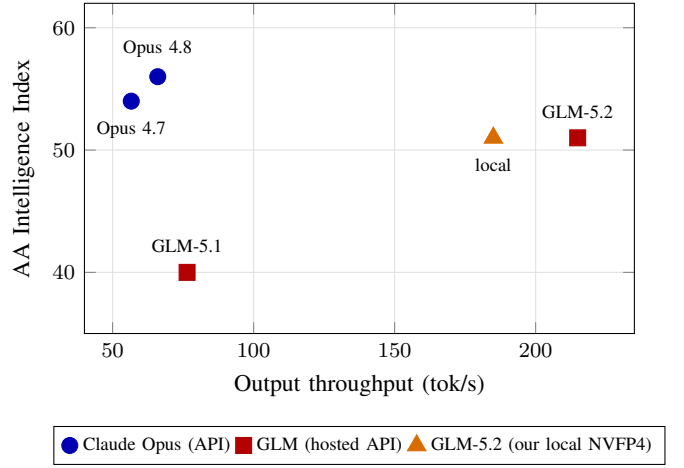
\begin{figure}[htbp]
\centering
\begin{tikzpicture}
\begin{axis}[
    xlabel={Output throughput (tok/s)},
    ylabel={AA Intelligence Index},
    xmin=40, xmax=235, ymin=35, ymax=62,
    width=\columnwidth, height=5.95cm,
    grid=both, grid style={line width=.1pt, draw=gray!15},
    major grid style={line width=.2pt, draw=gray!25},
    legend style={at={(0.5,-0.28)}, anchor=north, legend columns=-1, font=\scriptsize},
    label style={font=\small}, tick label style={font=\footnotesize},
]
\addplot[only marks, mark=*, blue!70!black, mark size=3pt] coordinates {(66.0,56) (56.6,54)};
\addlegendentry{Claude Opus (API)}
\addplot[only marks, mark=square*, red!70!black, mark size=3pt] coordinates {(214.9,51) (76.4,40)};
\addlegendentry{GLM (hosted API)}
\addplot[only marks, mark=triangle*, orange!85!black, mark size=4pt] coordinates {(185,51)};
\addlegendentry{GLM-5.2 (our local NVFP4)}
\node[anchor=south, yshift=4pt, font=\scriptsize] at (axis cs:66,56) {Opus 4.8};
\node[anchor=north, yshift=-4pt, font=\scriptsize] at (axis cs:56.6,54) {Opus 4.7};
\node[anchor=south, yshift=4pt, font=\scriptsize] at (axis cs:214.9,51) {GLM-5.2};
\node[anchor=south, yshift=4pt, font=\scriptsize] at (axis cs:76.4,40) {GLM-5.1};
\node[anchor=north, yshift=-4pt, font=\scriptsize] at (axis cs:185,51) {local};
\end{axis}
\end{tikzpicture}
\caption{Speed--quality trade-off (Artificial Analysis hosted-endpoint measurements, July 2026): output throughput versus Intelligence Index. The Claude models occupy the high-quality/low-speed upper-left; the GLM models occupy the high-speed/lower-quality lower-right. The orange triangle marks our local NVFP4 GLM-5.2 deployment ($\approx$185 tok/s, a single-tenant deployment observation).}
\label{fig:performance_chart}
\end{figure}

To provide context for the empirical productivity results, Fig.~\ref{fig:performance_chart} contrasts output throughput with the Artificial Analysis Intelligence Index for the four models, as measured on the vendors' hosted endpoints in July 2026~\cite{artificialanalysis2026}. There is a clear trade-off between throughput and reasoning quality: the Claude Opus 4.8 and 4.7 APIs lead on intelligence (Index 56 and 54) but deliver 66.0 and 56.6 output tok/s, whereas hosted GLM-5.2 sustains 214.9 tok/s at Index 51 (GLM-5.1: 76.4 tok/s at Index 40).

Our local NVFP4 deployment of GLM-5.2 on the shared NVL72 cluster sustained observed output speeds of roughly 185 tok/s under interactive agent load---a single-tenant deployment observation rather than a controlled benchmark, but of the same order as the hosted 214.9 tok/s figure and consistent with Blackwell's native 4-bit hardware acceleration, and approximately $2.8\times$ the Opus 4.8 API. Faster raw generation, however, is not expected to translate into faster end-to-end task completion in Period B: the elevated defect rate multiplied the number of repair iterations per task (Section~V-B), so review-and-repair overhead---not raw token throughput---dominates wall-clock time. We did not instrument wall-clock completion time directly; consistent with this expectation, commit cadence \textit{slowed} in Period B (median inter-commit interval 12.7 vs.\ 5.9 min, Section~V-G)---echoing the slowdown measured in the agentic-assistant field RCT~\cite{metr2025rct}---though that proxy also absorbs the higher repair volume.

\subsection{Qualitative Case Studies and Code Anomalies}
To understand the nature of coding errors generated by the on-premise quantized GLM-5.1/5.2 configuration, we examine two representative defects drawn from Period B fix commits (code paraphrased and identifiers genericized).

\subsubsection{Static Typing Violation (mypy)}
Listing~1 shows a type-discipline failure. The agent implemented an internal API call whose backend endpoint returns a \texttt{dict} for POST requests but a \texttt{list} for GET requests, and bound the result without the required union annotation. The un-annotated assignment fails \texttt{mypy} on the exception-handling branch, and a dedicated repair commit adding the \texttt{dict | list} annotation was required.

{\footnotesize\noindent\textbf{Listing 1.} GLM omits the union type (POST returns \texttt{dict}, GET returns \texttt{list}):
\begin{verbatim}
result = await client.request(token, payload)
# mypy: incompatible assignment on the
# except-branch. Fix:
result: dict | list = \
    await client.request(token, payload)
\end{verbatim}
}

\subsubsection{Environment-Behavior API Misuse (httpx proxy inheritance)}
In the second case, the agent constructed an \texttt{httpx} client for a cluster-internal service without disabling environment-variable trust. The client silently inherited the pod's \texttt{HTTP\_PROXY} settings, routing in-cluster traffic through an external proxy and breaking the integration---a defect invisible at compile time that surfaced only at runtime inside the Kubernetes network.

{\footnotesize\noindent\textbf{Listing 2.} GLM output inherits the proxy environment:
\begin{verbatim}
client = httpx.AsyncClient(base_url=URL)
# -> in-cluster calls routed via HTTP_PROXY
# Fix:
client = httpx.AsyncClient(
    base_url=URL,
    trust_env=False)  # bypass proxy for
                      # cluster-internal calls
\end{verbatim}
}

Both anomalies illustrate the pattern quantified in Table~\ref{tab:defects}: the local configuration's failures concentrate in typing discipline and in the correct use of API and environment semantics---categories where the frontier API model rarely required dedicated repair commits.

\subsection{Objective Workload and Developer-Experience Indicators}
Rather than rely on subjective self-report to characterize the developer's day-to-day experience, we derive a set of \textit{objective behavioral indicators} directly from the timestamped Git history. Each indicator is computed deterministically from commit metadata and is fully reproducible; together they operationalize the workload dimensions that instruments such as the NASA-TLX~\cite{hart1988nasatlx} capture subjectively, but without recall bias or self-report subjectivity. We define: (i) the \textit{debugging-spiral share}, the fraction of commits occurring inside a maximal run of three or more consecutive fix commits; (ii) the \textit{longest uninterrupted repair run}; (iii) the \textit{median inter-commit interval} within a working stretch (gaps under 8 hours), a proxy for the cadence of visible progress; (iv) the \textit{median daily active span}, the wall-clock hours between the first and last commit on active days, a proxy for the workflow's temporal footprint (sessions frequently ran unattended) rather than continuous human presence; (v) the \textit{mean rework multiplicity}, the average number of distinct commits touching each file. Table~\ref{tab:behavioral} reports these indicators for both periods.

\begin{table}[htbp]
\caption{Objective Behavioral Workload Indicators (from timestamped Git history)}
\label{tab:behavioral}
\begin{center}
\footnotesize
\setlength{\tabcolsep}{2.5pt}
\begin{tabular}{lrrr}
\toprule
\textbf{Indicator} & \textbf{A (Opus)} & \textbf{B (GLM)} & \textbf{B/A} \\
\midrule
Fix-commit share ($\FCR$) & 45.9\% & 74.9\% & $1.6\times$ \\
Commits in debugging spirals ($\ge 3$) & 35.0\% & 69.8\% & $2.0\times$ \\
Longest uninterrupted repair run & 18 & 58 & $3.2\times$ \\
Median inter-commit interval & 5.9 min & 12.7 min & $2.2\times$ \\
Median daily active span & 6.6 h & 12.4 h & $1.9\times$ \\
Mean rework (commits per file) & 1.61 & 2.32 & $1.4\times$ \\
\bottomrule
\end{tabular}
\end{center}
\end{table}

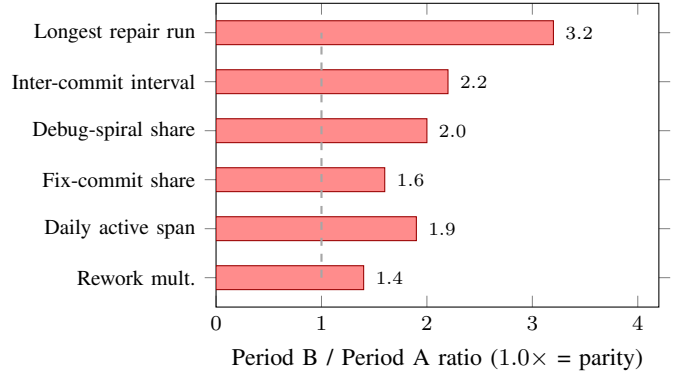
\begin{figure}[htbp]
\centering
\begin{tikzpicture}
\begin{axis}[
    xbar,
    bar width=9pt,
    width=0.84\columnwidth, height=5.6cm,
    xmin=0, xmax=4.2,
    xlabel={Period B / Period A ratio ($1.0\times$ = parity)},
    symbolic y coords={Rework,Span,FixShare,Spiral,Interval,LongRun},
    ytick=data,
    yticklabels={Rework mult., Daily active span, Fix-commit share, Debug-spiral share, Inter-commit interval, Longest repair run},
    nodes near coords, nodes near coords style={font=\scriptsize},
    every node near coord/.append style={/pgf/number format/precision=1, /pgf/number format/fixed, /pgf/number format/fixed zerofill, xshift=1pt},
    y tick label style={font=\footnotesize},
    x tick label style={font=\footnotesize},
    xlabel style={font=\small},
    enlarge y limits=0.12,
]
\addplot[fill=red!45, draw=red!60!black] coordinates {
    (1.4,Rework) (1.9,Span) (1.6,FixShare) (2.0,Spiral) (2.2,Interval) (3.2,LongRun)
};
\draw[dashed, gray!70, thick] (axis cs:1,Rework) -- (axis cs:1,LongRun);
\end{axis}
\end{tikzpicture}
\caption{Objective behavioral workload indicators expressed as the Period-B-to-Period-A ratio (Table~\ref{tab:behavioral}). All six axes exceed the $1.0\times$ parity line (dashed), i.e., every indicator worsens under the local GLM configuration; the repair-loop measures (longest run, inter-commit interval, spiral share) degrade most.}
\label{fig:behavioral}
\end{figure}

The indicators consistently show a heavier, more repair-dominated workload under the local GLM configuration: 69.8\% of Period B commits fell inside consecutive-fix runs (versus 35.0\% in Period A), the longest such run growing from 18 to 58 repair commits, with progress cadence, active-day span, and rework multiplicity all degrading in step (Table~\ref{tab:behavioral}; Fig.~\ref{fig:behavioral} plots each indicator's Period-B-to-Period-A ratio, every axis worsening with no offsetting improvement). These behavioral signals corroborate the quality findings of Section~V-B along a complementary axis: the cadence and span indicators derive purely from timestamps, and the run-structure indicators add temporal patterning to the keyword classification. Unlike a subjective questionnaire, they are free of recall bias and self-report framing---though not of a hypothesis-aware subject's labeling behavior, bounded by the invariance arguments of Section~VII-C.

\subsection{Hardware Serving Configuration and Quantization}
The local serving configuration follows the deployed Hugging Face checkpoint configurations~\cite{hfglm51,hfglm52}: both generations were served via the vLLM engine with PagedAttention~\cite{kwon2023vllm} on a shared NVIDIA GB200 NVL72 cluster. The smallest allocation on which the NVFP4 checkpoint can be served---a 4$\times$ B200 slice, one GB200 NVL72 compute tray---is the cost-accounting basis rather than the full physical cluster.

The arithmetic of 4-bit compression is what makes the one-tray footprint possible: a 753B-parameter model occupies roughly 1.5~TB in BF16---beyond even an eight-GPU allocation ($\sim$1.5~TB HBM) once KV cache and activations are included---whereas the NVFP4 checkpoint stores quantized weights at 4.5 bits per parameter (4-bit values plus one FP8 scale per 16-element block); with embeddings, router, and normalization tensors retained at higher precision, the deployed checkpoint measures $\approx$465~GB, fitting a 4$\times$ B200 tray ($\sim$750~GB HBM) with an FP8 KV cache. As noted in Section~II-C, vendor calibration reports claim near-lossless NVFP4 accuracy on standard benchmarks; Blackwell's native FP4 tensor-core path underlies the high observed generation throughput. This one-tray footprint is what enables the low amortized cost of \$2,560.00 USD in our economic model.

We emphasize a scoping limitation: we did not deploy FP16 or FP8 baselines of GLM-5.2 for a controlled quantization ablation (an FP16 baseline alone would occupy more than a full eight-GPU allocation). Consequently, our defect analysis cannot empirically separate quantization-induced degradation from the base model's intrinsic capability gap relative to Claude Opus. We instead treat the NVFP4 checkpoint as the unit of analysis---it is the artifact an enterprise would actually deploy---and flag the attribution question as a threat to construct validity in Section~VII-C.

\section{Discussion}

Before interpreting these results, we state their causal scope. The gap we measure---most sharply the $\approx$$3.6\times$ difference in defect-repair odds---is the \textit{joint} effect of a deployable bundle: base-model capability, NVFP4 quantization, vLLM serving, the agent harness (Claude Code vs.\ Opencode), and the fixed period ordering of a naturalistic study. It cannot be attributed to any single factor---least of all ``the model'' in isolation---so every claim below should be read at the level of the deployable \textit{configuration} an enterprise would adopt; factor decomposition is deferred to Section~VII-C.

To synthesize the empirical findings, Table~\ref{tab:overall} presents a multi-dimensional comparative summary of the commercial public API model and the on-premise open-weights cluster configuration.

\begin{table*}[htbp]
\caption{Overall Comparative Summary of Evaluated Paradigms}
\label{tab:overall}
\begin{center}
\footnotesize
\setlength{\tabcolsep}{3pt}
\begin{tabular}{lll}
\toprule
\textbf{Evaluation Dimension} & \textbf{Period A: API-based Opus} & \textbf{Period B: On-Premise GLM on NVL72} \\
\midrule
\textbf{Inference Infrastructure} & Public Cloud (Claude API) & Private Cluster (4$\times$ B200 tray, shared Blackwell) \\
\textbf{Quantization Format} & Vendor-managed (undisclosed) & NVFP4 (NVIDIA 4-Bit Floating Point) \\
\textbf{Serving Framework} & Vendor Hosted & Local vLLM Engine with PagedAttention \\
\textbf{Prompt Caching} & Enabled (Anthropic KV Cache, 99.3\% Hit Rate) & Enabled (vLLM Prefix Cache, 86.4\% Hit Rate) \\
\midrule
\textbf{Inference Speed (Output)} & 56.6--66.0 tok/s (hosted, Opus 4.7/4.8) & $\approx 185$ tok/s (single-tenant; $2.8\times$ vs.\ Opus 4.8) \\
\textbf{Model Capability (AA Index / SWE-bench Pro)} & 56 / 69.2\% & 51 / 62.1\% \\
\textbf{Task Development Protocol} & Specification-Driven Development (SDD) & Specification-Driven Development (SDD) \\
\midrule
\textbf{Git Commit Yield (non-merge)} & 246 Commits / 129,556 Insertions & 367 Commits / 93,281 Insertions \\
\textbf{Fix Commit Ratio (FCR)} & \textbf{45.93\%} (Low defect-repair overhead) & \textbf{74.93\%} (High defect-repair overhead) \\
\textbf{Syntax/Type Repair Share} & Low (1.8\% of fix commits) & Elevated (7.3\% of fix commits) \\
\midrule
\textbf{28-Day Compute / Infra Cost} & \$8,785.21 (Token-based API cost) & Shared \$3,453 / dedicated \$11,645 (incl.\ \$893 ops) \\
\textbf{Data Compliance / Governance} & Requires sending code to third-party endpoints & On-premise (no third-party transmission) \\
\bottomrule
\end{tabular}
\end{center}
\end{table*}

\subsection{The Cost--Quality Trade-off and Developer Churn}
Our results highlight a classic software quality trade-off. While the local infrastructure reduces direct compute expenditure by up to 71\% (shared allocation), it introduces substantial developer churn: the high $\FCR$ (74.93\%) of the GLM-5.1/5.2 configuration implies that the developer spent considerable work hours correcting suboptimal AI outputs.

We can model the total cost of development ($TC$) as:
\begin{equation}
TC = C_{\text{compute}} + C_{\text{ops}} + (T_{\text{dev}} \times R_{\text{dev}}).
\end{equation}
To parameterize this model under realistic industrial conditions in Taipei, Taiwan (all conversions at 32 NTD/USD), we set the developer wage rate to $R_{\text{dev}} = \$35.00$ USD per hour (NT\$1,120/hour). This approximates the \textit{total employer compensation cost} of a senior software engineer in Taiwan---official earnings statistics and market data (cf.~\cite{dgbas2026}) place senior engineer base pay near NT\$90k--130k per month, which grosses to roughly NT\$170k--190k per month after statutory labor/health insurance, pension contributions, and bonus provisions, i.e., about NT\$1,120 per working hour.

For the cloud API configuration (Period A), $C_{\text{ops}} = 0$ as infrastructure maintenance, electricity, and cooling are fully absorbed by the provider. For the private on-premise B200 allocation (Period B), we define $C_{\text{ops}}$ from Taiwan operating parameters:
\begin{enumerate}
    \item \textbf{Power and Cooling}: A 4$\times$ B200 serving tray draws approximately 6.0~kW under load, accounting for the roughly 1.2~kW per-GPU board power plus the amortized host system share. Over a 28-day cycle (672 hours) at Taipower's high-voltage industrial summer rate of approximately NT\$4.3 per kWh~\cite{taipower2025} (\$0.134 USD/kWh), direct electricity is \$542; applying a Power Usage Effectiveness (PUE) multiplier of 1.5 for sub-tropical cooling overhead yields a total power cost of $\approx\$813$; inference-time carbon emissions~\cite{luccioni2023bloom} are a further dimension our dollar-denominated model does not price.
    \item \textbf{MLOps Personnel Amortization}: A local site-reliability or MLOps engineer at a monthly employer cost of NT\$100,000 (\$3,125 USD/month), amortized across a shared cohort of ${\sim}100$ developers, contributes $\approx\$30$ USD per developer.
    \item \textbf{Facilities and Maintenance}: Space and data center bandwidth amortized at $\$50.00$ USD per developer.
\end{enumerate}
This yields a total operational overhead of $C_{\text{ops}} \approx \$893$ USD (NT\$28,576) per developer for Period B. The three components deliberately use different bases---power at the full-tray 672-hour draw, MLOps and facilities amortized across a 100-developer cohort. Charging the whole tray's power to one developer over-attributes idle-time energy to Period B, and adding local ops on top of a rental rate that already embeds facility overheads double-counts in the same direction; both are \textit{conservative} against our savings conclusion, so the true saving is if anything larger. We evaluate the developer's defect-repair time ($T_{\text{dev}}$) as the number of fix commits times a per-fix review time. This directly instantiates the developer-time term of the feedback-loop model (Section~IV-D): the timestamped commit history measures realized repair effort directly, so the $T_{\text{iter}}$ decomposition serves as the conceptual frame while the observed fix-commit volume supplies the quantity. Rather than assume an asymmetric per-fix duration, we apply a single \textit{symmetric} review time of 15 minutes (0.25 hours) to both configurations, so that the labor difference is driven entirely by the objectively counted difference in fix-commit volume rather than by a subjective per-fix penalty. The symmetric value is empirically grounded: the median inter-commit interval \textit{immediately preceding a fix commit} is nearly identical across periods (14.1 min in Period A, 13.2 min in Period B), so the per-fix repair cadence itself shows no asymmetry---the overall cadence gap (5.9 vs.\ 12.7 min, Section~V-G) stems from the surrounding non-repair workflow. Initial feature-development labor, common to both periods and of comparable magnitude, cancels in this between-configuration comparison, so the ``True TCO'' below is compute plus \textit{differential} repair labor rather than an absolute lifecycle cost.
\begin{itemize}
    \item \textbf{Period A (Opus)}: 113 fix commits $\times$ 0.25 h $\times$ \$35 $= \$988.75$ ($T_{\text{dev}}^A = 28.25$ h). The True TCO (Compute + Labor) is $\$9{,}773.96$ USD.
    \item \textbf{Period B (GLM)}: 275 fix commits $\times$ 0.25 h $\times$ \$35 $= \$2{,}406.25$ ($T_{\text{dev}}^B = 68.75$ h). Adding $C_{\text{ops}}$, the True TCO under shared allocation is $\$2{,}560.00 + \$893.00 + \$2{,}406.25 = \$5{,}859.25$ USD (NT\$187,496); under dedicated reservation it is $\$10{,}752.00 + \$893.00 + \$2{,}406.25 = \$14{,}051.25$ USD.
\end{itemize}

We caution that commit cadence is an upper-bound proxy for developer effort (it includes agent and test-runner wait time); we therefore report TCO to indicate magnitude, not precision. Comparing the two periods, the allocation regime decides the outcome: under shared allocation, local deployment saves 40.1\% of True TCO (\$5,859.25 vs.\ \$9,773.96, a net \$3,914.71 saving per 28 days), whereas under dedicated reservation it costs 43.8\% \textit{more} than the cached API (\$14,051.25 vs.\ \$9,773.96)---the 512 of 672 tray-hours billed beyond the standard seat erase the unit-cost advantage. In the shared regime, the dominant driver is that the cached-API compute bill (\$8,785) still dwarfs the tray amortization (\$2,560) plus the incremental repair labor (\$2,406 vs.\ \$989), so even a $2.4\times$ defect-repair burden does not erase the compute advantage. Instantiating the break-even criterion of Section~IV-D: $\Delta C_{\text{compute}} = \$8{,}785 - \$3{,}453 = \$5{,}332$ against $R_{\text{dev}} \times \Delta T_{\text{dev}} = \$35 \times 40.5\,\text{h} = \$1{,}418$---the shared-allocation advantage clears the criterion by $3.8\times$. The genuine cost of the local configuration is therefore not dollars but the developer-experience burden quantified objectively in Section~V-G.

\textit{Parameter sensitivity.} Because the cost model rests on several point estimates, we sweep the three most consequential: the wage rate (\$27--45/hr), the GPU lease rate (\$3--6 per GPU-hour), and the per-fix review time (10--25 min). The \textit{shared-allocation} conclusion is robust---on-premise True TCO savings stay positive across the entire sweep, ranging from about 9\% (all-adverse corner) to 56\% (all-favorable). The \textit{dedicated-reservation} option, by contrast, is dominated throughout: with the tray reserved around the clock, its True TCO exceeds the cached API's at every point of the sweep (from roughly 10\% above in the most favorable corner to more than double in the most adverse), so dedicated reservation only pays off when the tray serves substantially more than a single developer's workload. Attribution hours are a fourth axis: at the point estimates, the shared-allocation saving survives any per-developer attribution up to $\approx$405 of the 672 tray-hours ($\approx$2.5 standard seats, 60\% of the period's wall clock); only beyond that does the on-premise True TCO exceed the cached API's. The qualitative ordering---shared on-premise cheapest, then the pure API, then dedicated on-premise---holds throughout the sweep. One axis is held fixed: the Claude bill uses public list rates, though Anthropic also offers negotiated volume and committed-use discounts~\cite{anthropicpricing2026}; a committed-use reduction of $\approx$45\% in the API compute bill would by itself equalize the shared-allocation TCO, so this advantage is scoped to list-price API economics.

\subsection{Information Security and Governance Context}
Beyond mitigating the third-party data-governance risk already noted (Section~III-E), local NVL72 serving keeps in-scope inference data inside the corporate network.

However, operating a shared local GPU cluster introduces internal multi-tenancy risk: proprietary code, schemas, and request contexts from multiple departments co-reside in shared GPU memory and host logs. Our infrastructure therefore layers hardware-level isolation (NVIDIA Multi-Instance GPU partitions with dedicated compute and memory outside the tensor-parallel serving trays), network and container isolation (dedicated Kubernetes namespaces under strict network policies, mutual TLS in transit), and token-based request isolation (role-based access control via developer-specific tokens). Because prefix caching is enabled, KV prefixes may be reused across requests, so cross-tenant isolation is enforced at these hardware, network, and access-control layers---maintaining data sovereignty not only against external third parties but also internally across multi-tenant development cohorts.

\subsection{Cost--Quality Frontier of a Hybrid Routing Gateway}
To evaluate the proposed hybrid routing gateway, we perform an offline counterfactual replay over the pooled set of 613 real non-merge commits from both periods, using the empirically measured per-tier fix rates of each configuration (Section~V-C): Claude 15.1\%/61.2\%/63.4\% and GLM 46.5\%/80.6\%/86.9\% for High/Medium/Low tiers respectively. Each routing policy assigns every difficulty tier to one backend; infrastructure cost is amortized per commit by dividing each backend's measured period total (API \$8,785.21; shared GPU + ops \$3,453) by the pooled 613-commit denominator---so routing 100\% of the pooled workload to a backend reproduces its measured bill---and repair labor is priced per fix commit at the symmetric per-fix review time of Section~VI-A. This assumes each backend's infrastructure cost scales linearly with commit count and that the pooled workload's footprint matches that backend's own period---first-order approximations, not measured quantities. The Pure Local GLM TCO reported here (\$7,388) is deliberately \textit{not} the Period B actual of Section~VI-A (\$5,859): the replay reprices the \textit{pooled} 613-commit workload---not Period B's own 367 commits---at each backend's per-tier fix rates, so that all four policies are scored on an identical workload and are therefore mutually comparable, at the cost of not being directly comparable to the single-period figures.

\begin{table*}[htbp]
\caption{Simulated Routing Policies on the Pooled Real Workload (613 commits; two-stage bootstrap 95\% CIs over 10,000 resamples)}
\label{tab:hybrid_simulation}
\begin{center}
\begin{tabular}{lrrrr}
\toprule
\textbf{Metric} & \textbf{Pure Claude API} & \textbf{Pure Local GLM} & \textbf{Hybrid (High$\rightarrow$Claude)} & \textbf{Hybrid (High+Med$\rightarrow$Claude)} \\
\midrule
Workload share routed to API & 100\% & 0\% & 28.1\% & 62.5\% \\
Infrastructure + Ops Cost & \$8,785 & \$3,453 & \$4,949 & \$6,785 \\
Repair Labor Cost & \$2,634 & \$3,935 & \$3,463 & \$3,105 \\
\textbf{True TCO} & \textbf{\$11,419 [11,043, 11,795]} & \textbf{\$7,388 [7,084, 7,679]} & \textbf{\$8,412 [8,125, 8,703]} & \textbf{\$9,890 [9,550, 10,232]} \\
TCO saving vs.\ Pure API & --- & 35.3\% & 26.3\% & 13.4\% \\
\midrule
\textbf{Simulated Overall FCR} & \textbf{49.1\% [42.1, 56.1]} & \textbf{73.4\% [67.7, 78.8]} & \textbf{64.6\% [59.2, 70.0]} & \textbf{57.9\% [51.5, 64.3]} \\
\bottomrule
\end{tabular}
\end{center}
\end{table*}

\begin{figure}[htbp]
\centering
\begin{tikzpicture}
\begin{axis}[
    ybar stacked,
    bar width=14pt,
    width=\columnwidth, height=5.85cm,
    ylabel={Total Cost (USD)},
    label style={font=\footnotesize},
    y tick label style={font=\footnotesize},
    symbolic x coords={Pure Claude, Pure GLM, Hybrid H, Hybrid H+M},
    xtick=data,
    ymin=0, ymax=12000,
    scaled y ticks=false,
    ytick={0,2000,4000,6000,8000,10000,12000},
    yticklabels={0,2k,4k,6k,8k,10k,12k},
    legend style={at={(0.5,-0.15)}, anchor=north, legend columns=-1, font=\scriptsize},
    ymajorgrids=true,
    x tick label style={font=\small},
]
\addplot[fill=black!30, draw=black!55] coordinates {(Pure Claude,8785.00) (Pure GLM,3453.00) (Hybrid H,4949.00) (Hybrid H+M,6785.00)};
\addplot[fill=orange!40, draw=orange!60!black] coordinates {(Pure Claude,2634.00) (Pure GLM,3935.00) (Hybrid H,3463.00) (Hybrid H+M,3105.00)};
\legend{Compute + Ops Cost, Defect-Repair Labor Cost}
\end{axis}
\end{tikzpicture}
\caption{Simulated True TCO breakdown on the pooled 613-commit workload (Table~\ref{tab:hybrid_simulation}), comparing compute/infrastructure cost against developer defect-repair labor cost under Taiwan wage parameters. All four policies are evaluated on the same workload so the bars are directly comparable; Hybrid H is the High$\rightarrow$Claude routing policy and Hybrid H+M the High+Med$\rightarrow$Claude policy.}
\label{fig:tco_breakdown}
\end{figure}
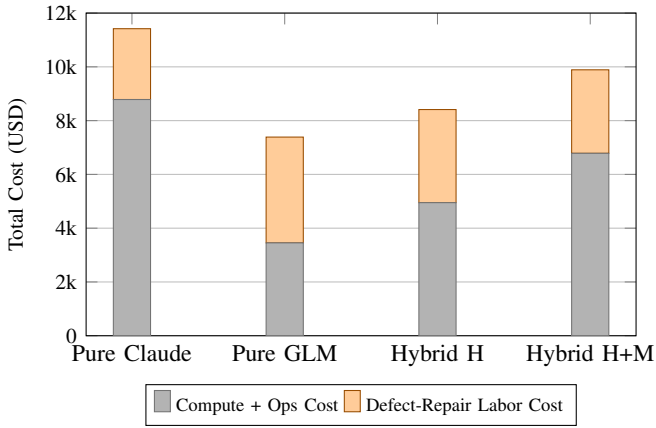

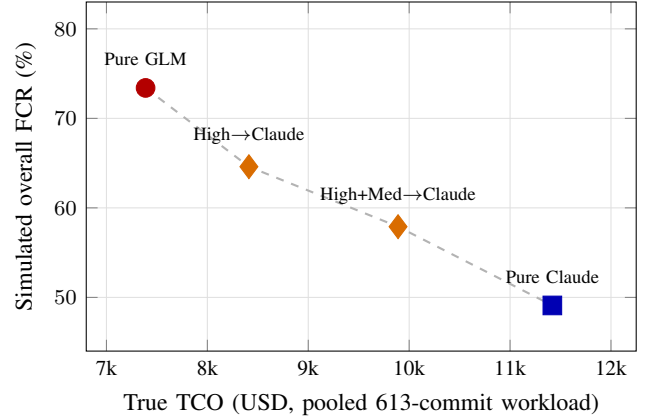
\begin{figure}[htbp]
\centering
\begin{tikzpicture}
\begin{axis}[
    xlabel={True TCO (USD, pooled 613-commit workload)},
    ylabel={Simulated overall FCR (\%)},
    xmin=6800, xmax=12250, ymin=44, ymax=83,
    width=\columnwidth, height=6.2cm,
    grid=both, grid style={line width=.1pt, draw=gray!15},
    major grid style={line width=.2pt, draw=gray!25},
    scaled x ticks=false,
    xtick={7000,8000,9000,10000,11000,12000},
    xticklabels={7k,8k,9k,10k,11k,12k},
    label style={font=\small}, tick label style={font=\footnotesize},
]
\addplot[thick, gray!60, dashed, mark=none, forget plot] coordinates {(7388,73.4) (8412,64.6) (9890,57.9) (11419,49.1)};
\addplot[only marks, mark=*, red!70!black, mark size=3.5pt] coordinates {(7388,73.4)};
\addplot[only marks, mark=diamond*, orange!85!black, mark size=4.5pt] coordinates {(8412,64.6) (9890,57.9)};
\addplot[only marks, mark=square*, blue!70!black, mark size=3.5pt] coordinates {(11419,49.1)};
\node[anchor=south, yshift=5pt, font=\scriptsize] at (axis cs:7388,73.4) {Pure GLM};
\node[anchor=south, yshift=5pt, font=\scriptsize] at (axis cs:8412,64.6) {High$\rightarrow$Claude};
\node[anchor=south, yshift=5pt, font=\scriptsize] at (axis cs:9890,57.9) {High+Med$\rightarrow$Claude};
\node[anchor=south, yshift=5pt, font=\scriptsize] at (axis cs:11419,49.1) {Pure Claude};
\end{axis}
\end{tikzpicture}
\caption{Cost--quality frontier of the four routing policies (Table~\ref{tab:hybrid_simulation}). The dashed line traces the Pareto frontier; shifting workload to the local model moves strictly left-and-up (lower TCO, higher FCR)---a deliberate trade-off, not a free lunch.}
\label{fig:frontier}
\end{figure}

Table~\ref{tab:hybrid_simulation} yields a sobering correction to the intuitive ``best of both worlds'' narrative. Because the local configuration's fix rate exceeds Opus's \textit{within every difficulty tier}---including 86.9\% on Low-difficulty commits---\textbf{no routing policy that sends any tier to the local model can match the pure-API defect profile}. Routing decisions therefore trace a monotonic cost--quality frontier rather than finding a dominating optimum: each increment of workload shifted to the local backend buys infrastructure savings at the price of a higher overall FCR and more repair labor (Fig.~\ref{fig:tco_breakdown} decomposes each policy's TCO into compute/infrastructure and repair-labor components). On this frontier, the High$\rightarrow$Claude hybrid is a defensible compromise (26.3\% TCO saving while protecting the highest-stakes work with the stronger model), and pure local remains the TCO minimum (35.3\% saving) for organizations that can absorb the defect churn. Notably, the High+Med hybrid still saves 13.4\% but delivers an FCR 9 points worse than pure Claude---illustrating that partial-quality strategies must be checked against the frontier, not assumed beneficial. A dynamic escalation policy (start locally, escalate on repeated self-healing failure) would require modeling escalation latency and retry token burn that our commit-level replay cannot resolve; we leave its evaluation to the gateway prototype in future work. Figure~\ref{fig:frontier} plots the four policies in cost--quality space, making the monotonic frontier explicit: moving left (cheaper) strictly moves up (worse FCR). Cost--quality routing has been explored on public benchmarks---cascades~\cite{chen2023frugalgpt} and learned routers~\cite{ong2025routellm}---and cost-controlled agent evaluation is an emerging norm~\cite{kapoor2024agents}; what this frontier adds is per-tier \textit{defect-repair} rates and repair labor measured on a production enterprise workload.

\subsubsection{Developer-Adaptive Thresholding (Design Sketch)}
The static routing policies above assign whole difficulty tiers uniformly, but the automation-bias risk of accepting a defective suggestion depends on the reviewing developer. We therefore sketch---as an explicitly \textit{unevaluated} design direction---an adaptive gateway that routes on the product of a normalized task complexity $C(T) \in [0, 1]$ (aggregating change size, module dependency fan-out, and target-file cyclomatic complexity) and a developer-experience coefficient $E(P) \in [0, 1]$ (estimated from the developer's historical Git-mined success rate): a task is routed to the local model only when $C(T)\,(1 - E(P)) \le \theta$ and to the cloud API otherwise, where $\theta$ is an enterprise safety tolerance. The intent is that harder tasks and less-experienced reviewers both bias routing toward the stronger cloud model, shrinking the low-yield manual-repair loops most likely to trap junior developers. We deliberately report no numeric instantiation: the present study does not vary developer profiles, so calibrating $\theta$, the complexity weighting, and $E(P)$ against measured per-developer outcomes requires the multi-developer replication we leave to future work.

\subsubsection{Uncertainty Quantification via Bootstrap Resampling}
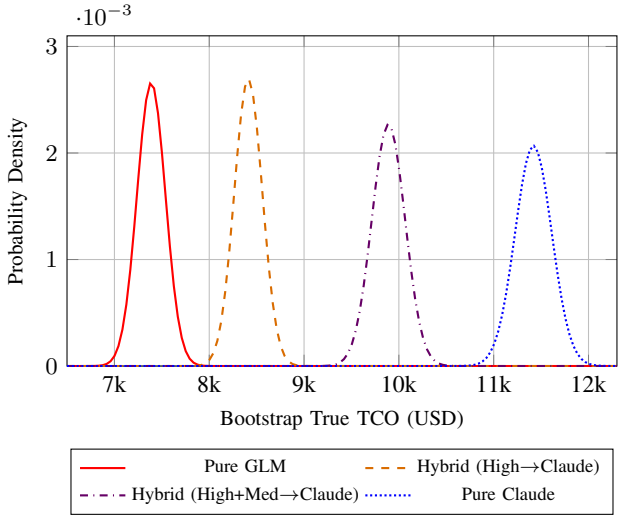
\begin{figure}[htbp]
\centering
\begin{tikzpicture}
\begin{axis}[
    xlabel={Bootstrap True TCO (USD)},
    ylabel={Probability Density},
    width=\columnwidth, height=5.95cm,
    label style={font=\footnotesize},
    xmin=6500, xmax=12300,
    ymin=0, ymax=0.0031,
    scaled x ticks=false,
    xtick={7000,8000,9000,10000,11000,12000},
    xticklabels={7k,8k,9k,10k,11k,12k},
    grid=both,
    x tick label style={font=\small},
    y tick label style={font=\small},
    legend style={font=\scriptsize, at={(0.5,-0.25)}, anchor=north, legend columns=2},
]
\addplot [red, domain=6500:12300, samples=140, thick] {1/(150*sqrt(2*pi))*exp(-((x-7388)^2)/(2*150^2))};
\addplot [orange!85!black, dashed, domain=6500:12300, samples=140, thick] {1/(148*sqrt(2*pi))*exp(-((x-8412)^2)/(2*148^2))};
\addplot [violet!80!black, dashdotted, domain=6500:12300, samples=140, thick] {1/(176*sqrt(2*pi))*exp(-((x-9890)^2)/(2*176^2))};
\addplot [blue, densely dotted, domain=6500:12300, samples=140, thick] {1/(193*sqrt(2*pi))*exp(-((x-11419)^2)/(2*193^2))};
\legend{Pure GLM, Hybrid (High$\rightarrow$Claude), Hybrid (High+Med$\rightarrow$Claude), Pure Claude}
\end{axis}
\end{tikzpicture}
\caption{Two-stage-bootstrap distributions (normal approximation of $10{,}000$ resamples) of True TCO for the four routing policies of Table~\ref{tab:hybrid_simulation}. The central 95\% intervals are pairwise disjoint; lower TCO trades against higher FCR.}
\label{fig:monte_carlo_pdf}
\end{figure}

Because both the per-commit outcomes and the per-cell fix rates are estimated from finite samples, presenting the simulated TCO and FCR as point estimates ignores sampling variance. We therefore quantify uncertainty by a \textit{two-stage} bootstrap: in each of $N = 10{,}000$ iterations, we first resample every backend$\times$tier cell at its observed period size to redraw its fix rate (propagating rate-estimation error), then resample the 613 pooled commits with replacement and redraw each commit's fix outcome as a Bernoulli trial from the redrawn rate of its tier and assigned backend. The resulting 95\% confidence intervals are reported in Table~\ref{tab:hybrid_simulation} and visualized in Fig.~\ref{fig:monte_carlo_pdf}. The four TCO intervals are pairwise disjoint, and the full TCO ordering (Pure GLM $<$ High$\rightarrow$Claude $<$ High+Med $<$ Pure Claude) persists in \textit{all} 10,000 paired resamples; the inverse FCR ordering persists in 99.8\% of resamples for the three-policy comparison (97.3\% across all four, whose adjacent FCR intervals overlap). The cost--quality frontier's ordering is thus statistically stable under workload and rate resampling---the trade-off itself, not merely its point estimate, is robust.

We stress that the bootstrap quantifies \textit{sampling} variance only. The replay rests on a stronger structural assumption: that each backend's per-tier fix rate is a transportable property of the (difficulty tier, backend) pair, independent of which tasks are routed. Because the two periods carried different task streams, this treats the tiers as exchangeable across periods---for instance, it assumes a High-difficulty task rerouted to Claude would inherit Claude's \textit{observed} High-tier rate rather than a rate specific to that task's content. The simulation is therefore a first-order projection, not a deployed-system measurement; a routing gateway can also change outcomes through mechanisms this static replay cannot capture (e.g., context handoff on escalation, or selection effects in which tasks reach which backend). We accordingly present the frontier as an illustrative decision aid whose \textit{ordering} is robust, and defer measured validation to the gateway prototype of Section~VIII (Future Work).

\subsection{Generalizability across Programming Language Paradigms}
\begin{figure}[htbp]
\centering
\begin{tikzpicture}
\begin{axis}[
    ybar,
    bar width=15pt,
    width=\columnwidth, height=5.65cm,
    ylabel={Fix Commit Ratio ($\FCR$, \%)},
    symbolic x coords={TypeScript (static), Python (dynamic)},
    xtick=data,
    ymin=0, ymax=100,
    ymajorgrids=true,
    nodes near coords, nodes near coords style={font=\scriptsize},
    every node near coord/.append style={/pgf/number format/precision=1, /pgf/number format/fixed, /pgf/number format/fixed zerofill},
    legend style={at={(0.5,-0.18)}, anchor=north, legend columns=-1, font=\footnotesize},
    tick label style={font=\footnotesize},
    label style={font=\small},
    enlarge x limits=0.5,
]
\addplot[fill=blue!40, draw=blue!60!black] coordinates {(TypeScript (static),41.7) (Python (dynamic),48.4)};
\addplot[fill=red!45, draw=red!60!black, postaction={pattern=north east lines, pattern color=red!60!black}] coordinates {(TypeScript (static),63.3) (Python (dynamic),86.8)};
\legend{Period A (Opus), Period B (GLM)}
\end{axis}
\end{tikzpicture}
\caption{Fix Commit Ratio by language paradigm. The Period-A$\rightarrow$B increase is far larger for dynamically typed Python ($48.4\%\rightarrow86.8\%$, $+38.4$ pts) than for statically typed TypeScript ($41.7\%\rightarrow63.3\%$, $+21.6$ pts): the compiler's static checks surface a share of the local model's errors inside the agent loop, before they become separate repair commits.}
\label{fig:language_fcr}
\end{figure}
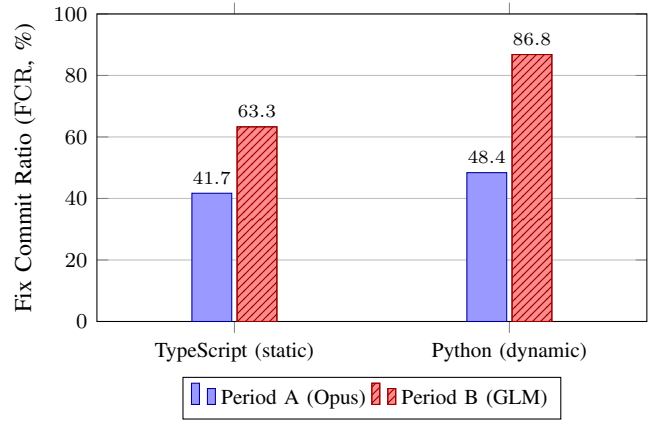

Our empirical evaluation was conducted within the production monorepo under study, a hybrid codebase mixing dynamically typed Python (37.6\% of Period A insertions) with statically typed TypeScript (17.1\%), plus specification documents and configuration (Table~\ref{tab:languages}). To analyze how the cost--quality trade-offs vary by language paradigm, we assign each non-merge commit a dominant language by the file types it touches, and compute per-language FCR for both periods:
\begin{itemize}
    \item \textbf{Statically Typed (TypeScript)}: In Period A (Opus), TypeScript-dominant work yielded 48 commits, of which 20 were fixes ($\FCR_{\text{TS}} = 41.7\%$). In Period B (GLM), 79 commits with 50 fixes ($\FCR_{\text{TS}} = 63.3\%$). TypeScript's compile-time checks surface interface and type violations immediately in the agent loop, keeping the repair cycle localized to fast compiler feedback.
    \item \textbf{Dynamically Typed (Python)}: In Period A, Python-dominant work yielded 95 commits with 46 fixes ($\FCR_{\text{Py}} = 48.4\%$). In Period B, 228 commits with 198 fixes ($\FCR_{\text{Py}} = 86.8\%$). Without compile-time type enforcement, incorrect interface assumptions in the local model's output tend to bypass initial static checks and surface only at runtime or in test suites (e.g., FastAPI dependency wiring and Pydantic validation), lengthening the repair feedback loop relative to compiler-caught defects.
\end{itemize}
In both periods, Python-dominant commits exhibit a higher FCR than TypeScript-dominant commits ($48.4\%$ vs.\ $41.7\%$ under Opus; $86.8\%$ vs.\ $63.3\%$ under GLM), and the deployment effect dwarfs the language effect: switching from Opus to the local GLM configuration sharply raises the per-language FCR in both paradigms (Fig.~\ref{fig:language_fcr}). This suggests our cost--quality findings are not an artifact of a single language ecosystem, while static typing shows a consistent directional defect-surfacing advantage (the within-period gaps are not individually significant)---in line with large-scale repository evidence associating type discipline with defect proneness~\cite{ray2014large}, an association whose magnitude subsequent replication work has questioned~\cite{berger2019impact}.

\section{Threats to Validity}
In this section, we discuss the potential threats to the validity of our empirical evaluation, categorized into internal, external, and construct validity.

\subsection{Internal Validity}
Internal threats concern factors that could affect the observed causal relationships. 
\begin{itemize}
    \item \textbf{Learning Curve (Maturation) Effect}: Since Period B (local GLM deployment) succeeded Period A (cloud API deployment), the developer's familiarity with the codebase naturally increased; any maturation effect would therefore \textit{favor} Period B. Within Period B, the FCR showed no significant decline over time---first half (days 1--14): 77.6\% (121/156), second half: 73.0\% (154/211); $\chi^2(1) = 1.00$, $p = 0.32$---indicating that four weeks of accumulated experience with the local model did not reduce its repair burden (Fig.~\ref{fig:weekly_fcr}). Within Period A, the FCR rose modestly from 42.2\% to 50.0\% ($p = 0.22$, not significant) as the work mix shifted from feature build-out toward stabilization and CI hardening in later weeks. Neither temporal pattern supports maturation as an explanation for the between-period gap: even Period A's most repair-heavy week (55.4\%) remains below Period B's \textit{least} repair-heavy week (63.8\%). Conversely, Period B inherited a codebase enlarged by Period A's $+$90.7k net lines, which could raise task hardness independently of the model; however, the repair premium persists within every change-size tier (Section~V-C), and the repair mix shifts toward \textit{local} syntax/type defects (Table~\ref{tab:defects})---a pattern repository growth alone does not explain.

\begin{figure}[htbp]
\centering
\begin{tikzpicture}
\begin{axis}[
    xlabel={Evaluation Week},
    ylabel={FCR (\%)},
    width=\columnwidth, height=6.05cm,
    tick label style={font=\footnotesize},
    label style={font=\footnotesize},
    xmin=0.5, xmax=8.5,
    ymin=0, ymax=100,
    xtick={1,2,3,4,5,6,7,8},
    ymajorgrids=true,
    grid style=dashed,
    legend style={at={(0.5,-0.32)}, anchor=north, legend columns=-1, font=\scriptsize},
    nodes near coords,
    every node near coord/.style={anchor=south, font=\scriptsize, color=black, /pgf/number format/fixed, /pgf/number format/precision=1, /pgf/number format/fixed zerofill}
]
\addplot[
    color=blue,
    mark=square*,
    thick,
    mark options={fill=blue!40},
    every node near coord/.append style={anchor={\ifnum\coordindex=1 north\else south\fi}}
] coordinates {
    (1,52.9) (2,35.1) (3,43.4) (4,55.4)
};
\addlegendentry{Period A (Opus)}

\addplot[
    color=red,
    mark=*,
    thick,
    mark options={fill=red!40},
    every node near coord/.append style={anchor={\ifcase\coordindex south\or north\or south\or north\fi}}
] coordinates {
    (5,80.6) (6,73.0) (7,87.7) (8,63.8)
};
\addlegendentry{Period B (GLM)}

\draw[black!60, thick, dotted] (axis cs:4,55.4) -- (axis cs:5,80.6);
\draw[dashed, black!60, thick] (axis cs:4.5,0) -- (axis cs:4.5,100);
\node[above, black, font=\scriptsize] at (axis cs:4.5,100) {Model Shift};
\end{axis}
\end{tikzpicture}
\caption{Weekly FCR trend across Period A (Weeks 1--4) and Period B (Weeks 5--8), computed from non-merge commits. Week 7 is a mid-period feature push and carries the highest weekly FCR (87.7\%, 81 commits). Every Period B week exceeds every Period A week by a wide margin.}
\label{fig:weekly_fcr}
\end{figure}
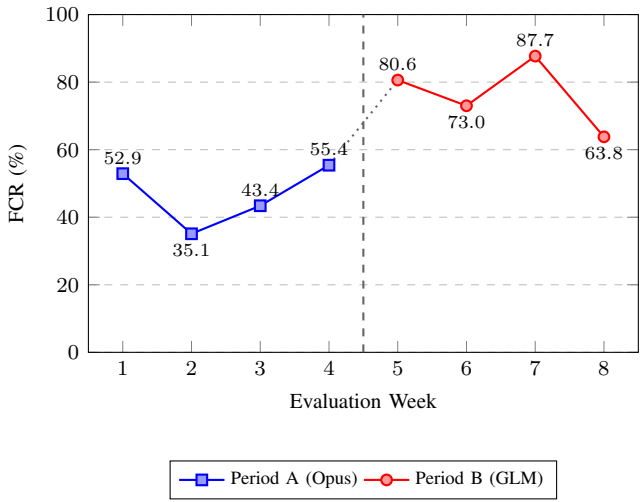
    \item \textbf{Task Selection Bias}: The difficulty imbalance runs \textit{opposite} to the confound (Section~IV-E), and the stratified $\mathrm{OR}_{\mathrm{MH}} = 3.61$ confirms the GLM premium within every tier (Section~V-C).
    \item \textbf{Experimenter and Demand-Characteristic Bias}: The sole subject is the hypothesis-aware first author, so the self-benchmark is exposed to expectancy effects. This is bounded structurally: the fix label is assigned by a deterministic keyword rule (Section~IV-C) rather than discretionary judgment, every counted commit passed the same enforced CI/review gate (Section~IV-G), and the premium holds within every difficulty tier (Section~V-C) and both language paradigms (Section~VI-D)---a structured pattern a diffuse expectancy effect would not produce.
\end{itemize}

\subsection{External Validity}
External threats concern the generalization of our findings.
\begin{itemize}
    \item \textbf{Single Project and Deployment Context}: Our empirical data derives from one production corporate AI PaaS monorepo under a single organizational deployment. While this controls for codebase and organizational noise, it may not immediately generalize to other codebases or larger, more heterogeneous engineering organizations. However, the scale and density of our data---over 44,000 LLM requests and 613 non-merge commits across 56 days---pin down the aggregate token, cost, and defect measures with high precision. This sharpens the aggregates' precision, not their generality; the inferential unit remains the day or the commit (Section~V-C). To facilitate broader generalization, a two-phase \textit{cross-over} protocol---randomly splitting a cohort of $N$ developers into two groups and swapping the API and on-premise configurations between phases---would neutralize inter-subject skill variance and linear learning curves.
    \item \textbf{Experience Level and Automation Bias}: The development activity under study reflects experienced-engineer usage patterns, with deep familiarity with the codebase. If replicated in a less-experienced context, the results may differ. Less-experienced developers may suffer from automation bias, blindly accepting incorrect code changes generated by the local GLM agent, which would raise the risk of defect leakage into the production branch. They may also experience a steeper surge in the behavioral workload indicators of Section~V-G (debugging-spiral share, active-hours span), owing to a reduced ability to resolve repetitive compile loops efficiently. Section~VI-C outlines an adaptive routing design intended to contain this risk by adjusting complexity thresholds to developer experience; it remains unevaluated, so the threat stands for less-experienced cohorts.
    \item \textbf{Model Specificity}: The evaluation is limited to the Claude and GLM model families. While these models represent current state-of-the-art hosted and local archetypes, rapid LLM iterations may alter the absolute performance scores.
\end{itemize}

\subsection{Construct Validity}
Construct threats concern the relationship between the theory and the observation.
\begin{itemize}
    \item \textbf{Fix Commit Ratio (FCR) Validity}: We measured quality via keyword mining of commit messages, which inherits a known limitation of label-based defect classification---misclassification can bias downstream analyses~\cite{herzig2013misclassification}. Trivial lint, typo, or formatting changes may be miscounted as repairs, and---as the taxonomy showed (Section~V-B)---a large share of Period A's fix commits were test/CI-infrastructure maintenance rather than model-generated-logic repairs. FCR is thus an upper bound on model-attributable defects for both periods, and the infrastructure share was larger in Period A (41.6\%), biasing the headline comparison \textit{against} our conclusion.

    \textit{Single-source and commit-granularity threats.} All quality signals derive from the same Git-commit stream; we have no independent defect oracle such as field-incident reports. Two properties bound this. First, both periods ran under an \textit{identical, enforced pre-push CI gate} (Table~\ref{tab:sdlc}), so trivial churn is blocked before push and the fix commits we count predominantly reflect defects that \textit{survived} a consistent gate; applied symmetrically, it cannot manufacture a between-period difference. Second, a change in commit \textit{granularity} alone cannot explain the pattern: the taxonomy shifts in \textit{composition} toward syntax/type and API defects under GLM (Table~\ref{tab:defects}), the premium persists \textit{within every difficulty tier} ($\mathrm{OR}_{\mathrm{MH}} = 3.61$), and it holds for both statically and dynamically typed code (Section~VI-D)---content-defined properties invariant to how repairs are packaged.
    \item \textbf{Benchmark Contamination}: Public base-model benchmark scores (e.g., SWE-bench) might suffer from pre-training data contamination; our study, conducted on a proprietary enterprise monorepo with no public internet exposure, provides a contamination-free complement to those leaderboards (Section~III-D).
    \item \textbf{Model Capacity vs.\ Agent Harness Confounding}: The observed differences are a combined product of base-model capacity, serving configuration (NVFP4 quantization), and harness design (Claude Code vs.\ Opencode). These cannot be fully separated in a naturalistic study: Claude Code is vendor-co-designed with Claude models, whereas Opencode adapts open-weights models through generic per-provider prompt profiles. The comparison should therefore be read at the level of \textit{deployable configurations}---the harness-plus-model bundles an enterprise would adopt---not as isolated base-model capability. Disentangling these factors (e.g., Claude through Opencode, or GLM through multiple harnesses) is left to controlled future experiments.
\end{itemize}

\section{Conclusion and Future Work}
We answer the four research questions of Section~I directly:
\begin{itemize}
    \item \textbf{RQ1 (Cost economics)}: Prompt caching (99.3\% hit rate) cuts the API's realized cost by 88.6\% to an effective \$0.57 per million tokens---below even the shared on-premise amortization (\$2.83/M). Per token, the cached API is the cheaper option at single-tenant on-premise utilization; its larger total bill stems only from its $16.9\times$ higher token volume.
    \item \textbf{RQ2 (Code quality)}: At comparable gross code churn, the local configuration was associated with a far higher defect-repair burden---$\FCR$ 74.9\% vs.\ 45.9\%, with defect-repair odds $2.6$--$4.9\times$ higher within every difficulty tier ($\mathrm{OR}_{\mathrm{MH}} = 3.61$).
    \item \textbf{RQ3 (True TCO)}: Shared on-premise allocation minimizes True TCO---a 40.1\% saving---whereas dedicated reservation costs 43.8\% more than the cached API; the shared saving is robust across the swept parameters at list-price API economics (Section~VI-A), while the dedicated option is dominated throughout.
    \item \textbf{RQ4 (Developer experience)}: The local configuration was associated with an objective developer-experience burden---roughly $2\times$ the debugging-spiral share, a $3.2\times$ longer worst-case repair run, and a $2.2\times$ slower commit cadence (a timestamp proxy that also absorbs unattended agent runtime).
\end{itemize}

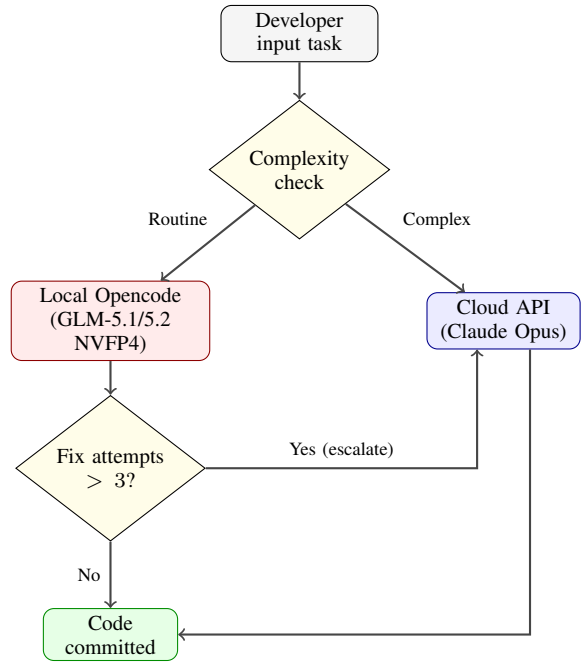
\begin{figure}[H]
\centering
\begin{tikzpicture}[
    block/.style={rectangle, draw, fill=gray!8, align=center, rounded corners, minimum height=0.62cm, inner sep=2.5pt, font=\footnotesize},
    decision/.style={diamond, draw, fill=yellow!10, align=center, inner sep=2pt, aspect=1.3, font=\footnotesize},
    line/.style={draw, ->, thick, gray!55!black},
    lab/.style={font=\scriptsize, black}
]
\node [block, text width=1.9cm] (req) at (0,0) {Developer input task};
\node [decision, text width=1.4cm] (dec1) at (0,-1.8) {Complexity\\check};
\node [block, text width=2.45cm, fill=red!8, draw=red!55!black] (glm) at (-2.5,-3.8) {Local Opencode\\(GLM-5.1/5.2 NVFP4)};
\node [block, text width=1.85cm, fill=blue!8, draw=blue!55!black] (claude) at (2.7,-3.8) {Cloud API\\(Claude Opus)};
\node [decision, text width=1.5cm] (dec2) at (-2.5,-5.75) {Fix attempts\\$>3$?};
\node [block, text width=1.6cm, fill=green!10, draw=green!55!black] (success) at (-2.5,-7.95) {Code committed};

\path [line] (req) -- (dec1);
\path [line] (dec1) -- node[pos=0.4, above left, lab] {Routine} (glm);
\path [line] (dec1) -- node[pos=0.4, above right, lab] {Complex} (claude);
\path [line] (glm) -- (dec2);
\path [line] (dec2) -- node[left, lab] {No} (success);
\path [line] (dec2) -| node[pos=0.25, above, lab] {Yes (escalate)} ([xshift=-0.35cm]claude.south);
\path [line] ([xshift=0.35cm]claude.south) |- (success);
\end{tikzpicture}
\caption{Proposed Tiered Routing Gateway: complex tasks route to the public API; routine tasks run locally, escalating after $>3$ failed fix attempts.}
\label{fig:routing}
\end{figure}

For future work, we propose the automated hybrid routing gateway of Fig.~\ref{fig:routing}: complexity pre-filtering routes high-stakes tasks to the cloud API; routine tasks start locally and escalate after a bounded number of failed self-healing iterations; evaluating its escalation latency and retry token burn is the primary open measurement question. We further plan parallel candidate generation with test-based filtering~\cite{chen2023codet}, exploiting local serving's zero marginal cost, and the cross-over, multi-developer replication of Section~VII-B.

\newpage
\section*{Acknowledgments}
The authors thank the PEGAVERSE platform engineering team for operating the shared NVIDIA GB200 NVL72 cluster and the enterprise quality-assurance and delivery pipeline spanning both evaluation periods.

\section*{Ethical Considerations}
This work is a self-study: the single developer analyzed is the first author, and the behavioral indicators (Section~V-G) derive from that author's own commit metadata, with consent. No customer or personal data was collected and no PII or proprietary source code is disclosed (the Section~V-F code is paraphrased); with no external human subjects, IRB oversight did not apply. The study targets deployable \textit{configurations}, not individuals---a point-in-time snapshot, not a durable model-family ranking.

\section*{Data Availability}
The proprietary enterprise codebase, raw telemetry, repository history, and analysis scripts remain confidential. Instead, the methodology is specified in full herein---telemetry queries, deterministic classification and difficulty rules, cross-branch \texttt{git patch-id} mining, and cost-model parameters (Sections~IV--VI)---sufficient to reproduce the pipeline on other data.

\bibliographystyle{IEEEtran}
\bibliography{references}

\begin{thebibliography}{10}
\providecommand{\url}[1]{#1}
\csname url@samestyle\endcsname
\providecommand{\newblock}{\relax}
\providecommand{\bibinfo}[2]{#2}
\providecommand{\BIBentrySTDinterwordspacing}{\spaceskip=0pt\relax}
\providecommand{\BIBentryALTinterwordstretchfactor}{4}
\providecommand{\BIBentryALTinterwordspacing}{\spaceskip=\fontdimen2\font plus
\BIBentryALTinterwordstretchfactor\fontdimen3\font minus
  \fontdimen4\font\relax}
\providecommand{\BIBforeignlanguage}[2]{{%
\expandafter\ifx\csname l@#1\endcsname\relax
\typeout{** WARNING: IEEEtran.bst: No hyphenation pattern has been}%
\typeout{** loaded for the language `#1'. Using the pattern for}%
\typeout{** the default language instead.}%
\else
\language=\csname l@#1\endcsname
\fi
#2}}
\providecommand{\BIBdecl}{\relax}
\BIBdecl

\bibitem{nvidia2024blackwell}
NVIDIA, ``{NVIDIA} {Blackwell} architecture: Technical brief,'' NVIDIA
  Corporation, Tech. Rep., 2024,
  \url{https://www.nvidia.com/en-us/data-center/technologies/blackwell-architecture/}.

\bibitem{yang2024sweagent}
J.~Yang, C.~E. Jimenez, A.~Wettig, K.~Lieret, S.~Yao, K.~Narasimhan, and
  O.~Press, ``{SWE-agent}: Agent-computer interfaces enable automated software
  engineering,'' in \emph{Adv. Neural Information Processing Systems
  (NeurIPS)}, 2024.

\bibitem{xia2024empirical}
C.~S. Xia, Y.~Deng, S.~Dunn, and L.~Zhang, ``Agentless: Demystifying
  {LLM}-based software engineering agents,'' \emph{arXiv preprint
  arXiv:2407.01489}, 2024.

\bibitem{jimenez2024swebench}
C.~E. Jimenez, J.~Yang, A.~Wettig, S.~Yao, K.~Pei, O.~Press, and K.~Narasimhan,
  ``{SWE-bench}: Can language models resolve real-world {GitHub} issues?'' in
  \emph{Proc. Int. Conf. Learning Representations (ICLR)}, 2024.

\bibitem{peng2023impact}
S.~Peng, E.~Kalliamvakou, P.~Cihon, and M.~Demirer, ``The impact of {AI} on
  developer productivity: Evidence from {GitHub} {Copilot},'' \emph{arXiv
  preprint arXiv:2302.06590}, 2023.

\bibitem{metr2025rct}
J.~Becker, N.~Rush, E.~Barnes, and D.~Rein, ``Measuring the impact of
  early-2025 {AI} on experienced open-source developer productivity,''
  \emph{arXiv preprint arXiv:2507.09089}, 2025.

\bibitem{kwon2023vllm}
W.~Kwon, Z.~Li, S.~Zhuang, Y.~Sheng, L.~Zheng, C.~H. Yu, J.~E. Gonzalez,
  H.~Zhang, and I.~Stoica, ``Efficient memory management for large language
  model serving with {PagedAttention},'' in \emph{Proc. 29th ACM Symp.
  Operating Systems Principles (SOSP)}, 2023, pp. 611--626.

\bibitem{yu2022orca}
G.-I. Yu, J.~S. Jeong, G.-W. Kim, S.~Kim, and B.-G. Chun, ``{Orca}: A
  distributed serving system for transformer-based generative models,'' in
  \emph{Proc. 16th USENIX Symp. Operating Systems Design and Implementation
  (OSDI)}, 2022.

\bibitem{lin2024awq}
J.~Lin, J.~Tang, H.~Tang, S.~Yang, W.-M. Chen, W.-C. Wang, G.~Xiao, X.~Dang,
  C.~Gan, and S.~Han, ``{AWQ}: Activation-aware weight quantization for
  on-device {LLM} compression and acceleration,'' in \emph{Proc. Machine
  Learning and Systems (MLSys)}, 2024.

\bibitem{frantar2023gptq}
E.~Frantar, S.~Ashkboos, T.~Hoefler, and D.~Alistarh, ``{GPTQ}: Accurate
  post-training quantization for generative pre-trained transformers,'' in
  \emph{Proc. Int. Conf. Learning Representations (ICLR)}, 2023.

\bibitem{xiao2023smoothquant}
G.~Xiao, J.~Lin, M.~Seznec, H.~Wu, J.~Demouth, and S.~Han, ``{SmoothQuant}:
  Accurate and efficient post-training quantization for large language
  models,'' in \emph{Proc. Int. Conf. Machine Learning (ICML)}, 2023.

\bibitem{hfglm52}
{NVIDIA and Zhipu AI}, ``{GLM-5.2-NVFP4} official model configuration,''
  \url{https://huggingface.co/nvidia/GLM-5.2-NVFP4}, 2026, accessed 2026-07-07.

\bibitem{giagnorio2025quantizing}
A.~Giagnorio, A.~Mastropaolo, S.~Afrin, M.~Di~Penta, and G.~Bavota,
  ``Evaluating the impact of post-training quantization on large language
  models for code generation,'' \emph{arXiv preprint arXiv:2503.07103}, 2025.

\bibitem{artificialanalysis2026}
{Artificial Analysis}, ``{Artificial Analysis}: {LLM} model intelligence,
  performance and price index,'' \url{https://artificialanalysis.ai/models},
  2026, accessed 2026-07-07.

\bibitem{morphllm2026}
{Morph}, ``{SWE-bench Pro} leaderboard (2026): Model scores,''
  \url{https://www.morphllm.com/swe-bench-pro}, 2026, accessed 2026-07-07.

\bibitem{openaipricing2026}
{OpenAI}, ``{GPT-5.5} model documentation,''
  \url{https://developers.openai.com/api/docs/models/gpt-5.5}, 2026, accessed
  2026-07-11.

\bibitem{anthropic2024promptcache}
Anthropic, ``Prompt caching: Speed up and lower the cost of your {API}
  requests,'' \url{https://claude.com/blog/prompt-caching}, 2024, accessed
  2026-07-07.

\bibitem{claudecode2026}
{Anthropic}, ``{Claude Code}: Agentic coding documentation,''
  \url{https://code.claude.com/docs}, 2026, accessed 2026-07-07.

\bibitem{opencode2026}
{Anomaly / SST}, ``{opencode}: {AI} coding agent built for the terminal,''
  \url{https://opencode.ai/docs/}, 2026, accessed 2026-07-07.

\bibitem{hfglm51}
{Zhipu AI and community quantization}, ``{GLM-5.1-NVFP4} model configuration,''
  \url{https://huggingface.co/lukealonso/GLM-5.1-NVFP4}, 2026, accessed
  2026-07-07.

\bibitem{langfuse2025telemetry}
{Langfuse Team}, ``Langfuse: Open source {LLM} engineering platform for
  telemetry and pricing audit,'' \url{https://langfuse.com}, 2025, accessed
  2026-07-07.

\bibitem{mockus2000identifying}
A.~Mockus and L.~G. Votta, ``Identifying reasons for software changes using
  historic databases,'' in \emph{Proc. Int. Conf. Software Maintenance (ICSM)},
  2000, pp. 120--130.

\bibitem{boehm1981software}
B.~W. Boehm, \emph{Software Engineering Economics}.\hskip 1em plus 0.5em minus
  0.4em\relax Prentice-Hall, 1981.

\bibitem{kitchenham2024recommendations}
B.~Kitchenham and L.~Madeyski, ``Recommendations for analysing and
  meta-analysing small sample size software engineering experiments,''
  \emph{Empirical Softw. Eng.}, vol.~29, no.~6, p. 137, 2024.

\bibitem{anthropicpricing2026}
{Anthropic}, ``Claude platform pricing,''
  \url{https://platform.claude.com/docs/en/about-claude/pricing}, 2026,
  accessed 2026-07-07.

\bibitem{b200pricing2026}
{GetDeploying}, ``{NVIDIA B200} cloud pricing comparison across providers,''
  \url{https://getdeploying.com/gpus/nvidia-b200}, 2026, accessed 2026-07-07.

\bibitem{hart1988nasatlx}
S.~G. Hart and L.~E. Staveland, ``Development of {NASA-TLX} ({Task Load
  Index}): Results of empirical and theoretical research,'' in \emph{Human
  Mental Workload}, P.~A. Hancock and N.~Meshkati, Eds.\hskip 1em plus 0.5em
  minus 0.4em\relax North-Holland, 1988, pp. 139--183.

\bibitem{dgbas2026}
{Directorate-General of Budget, Accounting and Statistics (DGBAS), Taiwan},
  ``Earnings and productivity statistics,''
  \url{https://earnings.dgbas.gov.tw/}, 2026, accessed 2026-07-07.

\bibitem{taipower2025}
{Taiwan Power Company}, ``Electricity tariff tables (effective 2025-10-01),''
  \url{https://www.taipower.com.tw/2289/2290/46940/}, 2025, accessed
  2026-07-07.

\bibitem{luccioni2023bloom}
A.~S. Luccioni, S.~Viguier, and A.-L. Ligozat, ``Estimating the carbon
  footprint of {BLOOM}, a 176{B} parameter language model,'' \emph{J. Mach.
  Learn. Res.}, vol.~24, 2023.

\bibitem{chen2023frugalgpt}
L.~Chen, M.~Zaharia, and J.~Zou, ``{FrugalGPT}: How to use large language
  models while reducing cost and improving performance,'' \emph{Trans. Mach.
  Learn. Res.}, 2024.

\bibitem{ong2025routellm}
I.~Ong, A.~Almahairi, V.~Wu, W.-L. Chiang, T.~Wu, J.~E. Gonzalez, M.~W. Kadous,
  and I.~Stoica, ``{RouteLLM}: Learning to route {LLMs} with preference data,''
  in \emph{Proc. Int. Conf. Learning Representations (ICLR)}, 2025.

\bibitem{kapoor2024agents}
S.~Kapoor, B.~Stroebl, Z.~S. Siegel, N.~Nadgir, and A.~Narayanan, ``{AI} agents
  that matter,'' \emph{arXiv preprint arXiv:2407.01502}, 2024.

\bibitem{ray2014large}
B.~Ray, D.~Posnett, V.~Filkov, and P.~Devanbu, ``A large scale study of
  programming languages and code quality in {GitHub},'' in \emph{Proc. 22nd ACM
  SIGSOFT Symp. Foundations of Software Engineering (FSE)}, 2014, pp. 155--165.

\bibitem{berger2019impact}
E.~D. Berger, C.~Hollenbeck, P.~Maj, O.~Vitek, and J.~Vitek, ``On the impact of
  programming languages on code quality: A reproduction study,'' \emph{ACM
  Trans. Program. Lang. Syst.}, vol.~41, no.~4, pp. 21:1--21:24, 2019.

\bibitem{herzig2013misclassification}
K.~Herzig, S.~Just, and A.~Zeller, ``It's not a bug, it's a feature: How
  misclassification impacts bug prediction,'' in \emph{Proc. 35th Int. Conf.
  Software Engineering (ICSE)}, 2013, pp. 392--401.

\bibitem{chen2023codet}
B.~Chen, F.~Zhang, A.~Nguyen, D.~Zan, Z.~Lin, J.-G. Lou, and W.~Chen,
  ``{CodeT}: Code generation with generated tests,'' in \emph{Proc. Int. Conf.
  Learning Representations (ICLR)}, 2023.

\end{thebibliography}

\end{document}